\documentclass[pdftex,twocolumn,epjc3]{svjour3}          

\RequirePackage[T1]{fontenc}

\smartqed  
\usepackage{amsmath}
\usepackage{indentfirst}
\usepackage{amssymb,dcolumn}
\usepackage{latexsym}
\usepackage{graphics}
\usepackage{setspace}
\usepackage{graphicx,float}
\usepackage[colorlinks,linkcolor=blue,citecolor=blue,urlcolor=blue,hyperindex]{hyperref}
\usepackage{color}
\usepackage{slashed}
\usepackage{CJKutf8}
\RequirePackage{graphicx}
\RequirePackage{mathptmx}      
\RequirePackage{flushend}
\RequirePackage[numbers,sort&compress]{natbib}
\RequirePackage[colorlinks,citecolor=blue,urlcolor=blue,linkcolor=blue]{hyperref}

\journalname{Eur. Phys. J. C}

\begin{document}
\begin{CJK}{UTF8}{<font>}
\title{QED and accretion flow models effect on optical appearance of Euler-Heisenberg black holes}

\author{Xiao-Xiong Zeng\thanksref{addr1,e1}
        \and
        Ke-Jian He\thanksref{addr2,e2}
        \and
        Guo-Ping Li\thanksref{addr3,e3}
        \and
        En-Wei Liang\thanksref{addr4,e4}
        \and
        Sen Guo\thanksref{addr4,e5}}
\thankstext{e1}{e-mail: xxzengphysics@163.com}
\thankstext{e2}{e-mail: kjhe94@163.com}
\thankstext{e3}{e-mail: gpliphys@yeah.net}
\thankstext{e4}{e-mail: lew@gxu.edu.cn}
\thankstext{e5}{Corresponding author, e-mail: sguophys@126.com}
\institute{\label{addr1} State Key Laboratory of Mountain Bridge and Tunnel Engineering, Department of Mechanics, Chongqing Jiaotong University, Chongqing 400074, People's Republic of China \and \label{addr2} Department of Physics and Chongqing Key Laboratory for Strongly Coupled Physics, Chongqing University, Chongqing 401331, People's Republic of China \and \label{addr3} School of Physics and Astronomy, China West Normal University, Nanchong 637000, People's Republic of China \and \label{addr4} Guangxi Key Laboratory for Relativistic Astrophysics, School of Physical Science and Technology, Guangxi University, Nanning 530004, People's Republic of China}

\date{Received: date / Accepted: date}

\maketitle

\begin{abstract}
Taking the quantum electrodynamics (QED) effect into account, we investigate the geometrical-optics appearance of the Euler-Heisenberg (EH) black hole (BH) under the different accretion flows context, which depends on the BH space-time structure and different sources of light. The more significant magnetic charge leads to the smaller shadow radius for the EH BH, while the different values of the EH parameter do not ruin it. Different features of the corresponding two-dimensional shadow images are derived for the three optically thin accretion flow models. It is shown that the total observed intensity in the static spherical accretion flow scenario leads than that of the infalling spherical accretion flow under same parameters, but the size and position of the EH BH shadows do not change in both of these accretions flows, implying that the BH shadow size depends on the geometric space-time and the shadows luminosities rely on the accretion flow morphology. Of particular interest is that a thin disk accretion model illuminated the BH, we found that the contribution of the lensing ring to the total observed flux is less than $5\%$, and the photon ring is less than $2\%$, indicating that the direct emission dominates the optical appearance of the EH BH. It is also believed that the optical appearance of the BH image depends on the accretion disk radiation position in this scenario, which can serve as a probe for the disk structure around the active galactic nucleus (AGN) of M87$^{*}$ like.
\end{abstract}

\section{Introduction}
\label{sec:intro}
\setlength{\parindent}{2em}
A black hole (BH) is the solution to Einstein's field equations of general relativity (GR) describing regions of space-time that have undergone gravitational collapse. The first convincing evidence to prove the existence of BHs in astronomical observations is the Laser-Interferometer Gravitational Wave-Observatory (LIGO) found that the gravitational wave emission from the coalescence of two BHs \cite{1}. Another big advance with the BH observations is the Event Horizon Telescope (EHT) collaboration debuted the near horizon image around the supermassive BH in the center of the Messier (M) 87$^{*}$ elliptical Galaxy ($M=6.5 \pm 0.7 \times 10^{9} M_{\odot}$). The BH image reports a compact asymmetric ring-like source morphology, which present a bright ring-shaped lump of radiation surrounding a circular dark region of an estimated 6.5 billion solar masses \cite{2,3,4,5,6,7}. It convincingly confirms the existence of BHs in our Universe. The EHT measured the corresponding linear polarimetric shadow image of the M87$^{*}$, crucial to understanding the launching of energetic jets from its core. It carries the information about the structure of the magnetic fields responsible for the synchrotron emission and found that magnetically arrested accretion disks surround the M87$^{*}$ (the mass accretion rate onto the BH of $(3-20) \times 10^{-4} M_{\odot} yr^{-1}$) \cite{8,9}. More excitingly, EHT obtained the first horizon-scale radio observations of the Sagittarius (Sgr) A$^{*}$ in our own Milky Way, which shows the measured ring size of the Sgr A$^{*}$ is consistent with the shadow critical curve predicted in GR within 10$\%$ \cite{eht-1,eht-2,eht-3,eht-4,eht-5,eht-6}.
\\
\indent
An astrophysical BH provides a constant space-time structure, but it can be illuminated by external sources of the luminous accretion material, leading the BH to have a variety of shapes and emit a variety of colors. The light from the accretion material is bent toward the singularity by the BH's gravitational field if the light passes close to a BH. This makes it possible to investigate the optical appearance of the BH from accretion flow. Possible observational characteristics of the BHs shadows surrounded by different accretion flows were studied for a long time. By establishing a thin-disk accretion flow model, Luminet presented that the emergence of the BH shadow and ring depends on the accretion flow position since the relative intrinsic intensity of radiation emitted by the disk accretion is a function of the BH radius \cite{10}. Falcke $et~al.$ created a ray-tracing code and proposed that the BH shadow is equivalent to the gravitational lensing effect by considering the radiation of a hot optically-thin accretion flow surrounding a supermassive BH in the center of our Galaxy \cite{11}. Assuming that the Schwarzschild BH is surrounded by a thin and heavy disk accretion flow, Cunha $et~al.$ investigated the BH gravitational lensing effect in this model and displayed an almost equatorial observer who could observe different patches of the sky near the equatorial plane \cite{12}. Narayan $et~al.$ studied the Schwarzschild BH shadow under a simple spherical accretion flow model, revealing that the optical appearance of the spherically accreting BH is independent of the inner radius at which the accreting gas stops radiating \cite{13}. By considering an optically/geometrically thin disk accretion flow model, Gralla $et~al.$ investigated the shadow and ring of the Schwarzschild BH shadow and found the image brightness diverges logarithmically at the BH photon ring \cite{14}.
\\
\indent
These experiments for the shadow image would be robust probes GR, and can further understands the properties of BHs and test the other modified theories. Considering the spherical accretion flow, Zeng $et~al.$ investigated the shadow of the Gauss-Bonnet BH and discussed the influence of the quintessence dark energy on the spherical BH. They found that the BH shadow depends on the physical properties of the accretion flow \cite{15,16}. In the framework of the Einstein-Gauss-Bonnet-Maxwell gravity, Ma $et~al.$ explored the photon sphere and shadow of the static spherically symmetric charged BH, which found that the shadow and photon sphere satisfy the sequence of inequalities to relate a BH's horizon and mass \cite{17}. Peng $et~al.$ argued that the BH shadow feature may provide observational evidence for the quantum effect of GR by investigating the quantum corrected Schwarzschild BH \cite{18}. Gao $et~al.$ studied the gravitational lensing effect of a hairy BH in Einstein-scalar-Gauss-Bonnet gravity and found that the BH shadow cast in this gravity is consistent with the measurement by the EHT, while other lensing observables are beyond the present capacity \cite{19}. Li $et~al.$ investigated the observational appearances of the global monopole BH illuminated by various accretions under the $f(R)$ gravity theory. They found that the BH shadows and the related rings with some different observable features can be used to distinguish BHs from different gravity theories \cite{20}. Guo $et~al.$ considered a charged BH surrounded by a perfect fluid radiation field within the framework of the Rastall gravity. They investigated the shadow and photon sphere of this BH with the static/infalling spherical accretion background and obtained that the shadow luminosity of this BH with infalling spherical accretion is dimmer than that of the static spherical accretion, but the photon sphere luminosity is brighter than the static one \cite{21}. Notably, they also showed that the luminosities of both the shadows and rings of the Hayward BH are affected by the accretion flow property and the BH magnetic charge \cite{22}. The BH shadow characteristics in the various gravity are also investigated in \cite{Guo,Gan,Guo-1,Wang,Guo-2,Walia,23,24,25}.
\\
\indent
One of the biggest problems in GR is the singularities that lay at the beginning of the Universe and also at the center of a BH. Maxwell's equations are known to exhibit singularities which cause the divergence problems in Maxwell's theory. Based on Dirac's positron theory, Euler and Heisenberg proposed a new approach to describe the electromagnetic field, taking into account the one-loop corrections to quantum electrodynamics (QED) and explaining the vacuum polarization in QED \cite{26}. Using this new method, the nonlinear electrodynamic models can explain the inflation of the universe in the early times \cite{27}. By considering the one-loop effective lagrangian density coupled with the Einstein field equation, Yajima $et~al.$ obtained the Euler-Heisenberg (EH) BH solution \cite{28}. Kruglov explored the vacuum birefringence effect of the nonlinear electromagnetic field and calculated the field effective Lagrangian, which found that the improvement in the PVLAS experiment would correct the relation between two parameters in the effective Lagrangian \cite{Kruglov}. Considering classical models of nonlinear electrodynamics with polynomial self-interaction of the electromagnetic field, Costa $et~al.$ proposed that the collation with QED results in the total field energy of a point elementary charge about twice the electron mass \cite{Costa}. By investigating the EH-type model of nonlinear electrodynamics with two parameters, Kruglov further obtained the charged black hole solution in the framework of nonlinear electrodynamics. It also obtained the corrections to Coulomb's law at $r \rightarrow \infty$ and studied the energy conditions \cite{Kruglov-1}. It is worth mentioning that Kruglov further derived the effective geometry induced by nonlinear electrodynamics corrections and determined the shadow's size of thee regular non-rotating magnetic BHs \cite{Kruglov-2}.
\\
\indent
Although the properties of BH in the nonlinear electrodynamic context have been extensively studied, the impact of the non-linear electrodynamics on the optical appearance of the EH BH shadow within the framework of different accretion flow models is an opening question. Meanwhile, it is unclear whether the accretion form affects the EH BH shadow optical appearance since the BH shadows exhibit various interesting observation characteristics under different accretion flow models. This paper focuses on this issue. The BH shadow is a powerful tool to investigate the observed characteristics of the BHs, it is interesting to investigate the optical appearance of the EH BH and know the effect of the QED effect on it. Moreover, to test the effects of the space-time structure on the EH BH shadow, observation effects can also be investigated with a varying magnetic charge. We investigate the optical appearance of the EH BH within three different optically thin accretion flows context and analysis the EH BH observation characteristics under varying magnetic charges by taking the QED effect into account.
\\
\indent
This paper is organized as follows. Section \ref{sec:2} briefly discuss the effective geometry of the EH BH and derives the light ray trajectory by using the ray-tracing method. In section \ref{sec:3}, we present the shadows and rings as well as the corresponding optical appearance based on the three accretion flow models. We draw the conclusions in section \ref{sec:4}.

\section{\textbf{EH BH the effective geometry and light deflection}}
\label{sec:2}
The four-dimensional action of the GR coupled with the non-linear electrodynamics can be described as \cite{Salazar}
\begin{equation}
\label{1-1}
S=\frac{1}{4\pi}\int {\rm d}^{2}x \sqrt{-g}\Big[\frac{1}{4}R-\mathcal{L}(\mathcal F, G)\Big],
\end{equation}
where $g$ is the metric tensor determinant, $R$ is the Ricci scalar. $\mathcal{L}(\mathcal{F},G)$ is the non-linear electrodynamics Lagrangian, depending on electromagnetic invariants, where $\mathcal{F}=\frac{1}{4}F_{\rm \mu \nu}F^{\rm \mu \nu}$, $G=\frac{1}{4}F_{\rm \mu \nu}^{*}F^{\rm \mu \nu }$ with $F_{\rm \mu \nu}$ denoting the electromagnetic field strength tensor in which $^{*}F^{\rm \mu \nu}=\epsilon_{\rm \mu \nu \sigma \rho}F^{\sigma \rho}/(2\sqrt{-g})$ its dual. The completely antisymmetric tensor is $\epsilon_{\rm \mu \nu \sigma \rho}$, which satisfies $\epsilon_{\rm \mu \nu \sigma \rho} \epsilon^{\rm \mu \nu \sigma \rho}=-4!$. The Lagrangian density of the EH non-linear electrodynamics is given by \cite{26}
\begin{equation}
\label{1-2}
\mathcal{L}(\mathcal{F},G)=-\mathcal{F}+\frac{a}{2}\mathcal{F}^{2}+\frac{7a}{8}G^{2},
\end{equation}
where $a=\frac{8\alpha^{2}}{45m^{4}}$ is the EH parameter, regulating the intensity of the non-linear electrodynamics contribution. $\alpha$ is the fine structure constant and $m$ is the electron mass, hence the EH parameter is written as $\alpha / E_{\rm c}^{2}$. For $a=0$, we recover the Maxwell electrodynamics, i.e. $\mathcal{L}=-\mathcal{F}$.

\par
In this paper, we focus on the static and spherically symmetric metric of the 4D compact objects, the line element is
\begin{equation}
\label{1-3}
{\rm d}s^{2}=-f(r){\rm d}t^{2}+f(r)^{-1}{\rm d}r^{2}+r^{2}{\rm d}\Omega^{2},
\end{equation}
where ${\rm d} \Omega^{2} = {\rm d} \theta^{2}+\sin^{2} \theta {\rm d} \phi^{2}$, $f(r)$ is the EH BH metric potential and it can be written as \cite{28},
\begin{equation}
\label{1-4}
f(r)=1-\frac{2M}{r}+\frac{Q^{2}}{r^{2}}-\frac{aQ^{4}}{20 r^{6}}.
\end{equation}
It was derived originally as an exact solution to the semi-classical Einstein equations with one-loop quantum corrections to QED. $M$ is the BH mass, $Q$ is the BH magnetic charge and the EH parameter $a$ satisfies $0 \leq a \leq \frac{32}{7}Q^{2}$ \cite{29}. For the $a \rightarrow 0$, the standard Maxwell electrodynamics is recovered, the EH BH will degenerate into the Reissner-Nordstr\"{o}m (RN) BH. Note that, Bret\'{o}n proposed a rotating BH solutions in the EH theory \cite{30}, we only consider the optical appearance of a static spherically symmetric EH BH solution in this analysis since the behavior of photons around the spin BH are rather differently from the static spherically symmetric BH.

\par
In order to investigate the light deflection near the EH BH, the movement of light ray in this space-time needs to be understood. Note that, the effective geometry induced by the EH non-linear electrodynamics effects, which correct the background geometry along the null geodesics of which photons would usually propagate \cite{Novello,Lorenci}. Taking this approach into account in our analysis, the photons propagate along the null geodesics of this effective geometry. The null geodesics of the photon paths are described by the effective space-time, that is \cite{Novello,Lorenci,Kruglov-2,31}
\begin{equation}
\label{1-5}
g_{\rm eff}^{\rm \mu \nu}=\mathcal{L}_{\mathcal F} g^{\rm \mu \nu}-\mathcal{L}_{\mathcal{F} \mathcal{F}}F^{\rm \mu}_{\rm \alpha}F^{\rm \alpha \nu},
\end{equation}
where $\mathcal{L}_{\mathcal{F}} \equiv \frac{\partial \mathcal{L}}{\partial \mathcal{F}}$. The effective geometry seen by photons on the background of the EH BH can be rewritten as
\begin{equation}
\label{1-6}
{\rm d}s_{\rm eff}^{2}=H(r)\big(-f(r){\rm d}t^{2}+f(r)^{-1}{\rm d}r^{2}\big)+h(r)r^{2}{\rm d}\Omega^{2},
\end{equation}
in which $H(r)=-\mathcal{L}_{\mathcal{F}}$ and $h(r)=-\mathcal{L}_{\mathcal{F}}-\mathcal{L}_{\mathcal{F}\mathcal{F}}\frac{Q^{2}}{r^{4}}$ \cite{Kruglov-2}. Therefore, we have
\begin{equation}
\label{1-7}
H(r)=1+\frac{a(aQ^{3}-2Qr^{4})^{2}}{8r^{12}},
\end{equation}
and
\begin{equation}
\label{1-8}
h(r)=1+\frac{a(aQ^{3}-2Qr^{4})^{2}}{8r^{12}}-\frac{aQ^{2}}{r^{4}}.
\end{equation}
Note that the function $H(r)$ and $h(r)$ must be positive in order that the effective geometry does not change its signature during the photon motion.

\par
Hence, the Lagrangian of the EH BH space-time is
\begin{equation}
\label{1-9}
\mathcal{L}= \frac{1}{2}\Big(H(r) \big(f(r)\dot{t}^{2} - f(r)^{-1} \dot{r}^{2}\big) - h(r) r^{2}(\dot{\theta}^{2}+\sin^{2}\theta \dot{\phi}^{2})\Big),
\end{equation}
where $\dot{x}\equiv {\rm d} x/{\rm d} \tau$, the $\tau$ is the affine parameter. Since we only consider the photons that move on the equatorial plane ($\theta=\pi/2$, $\dot{\theta}=0$ and $\ddot{\theta}=0$), and the EH BH Lagrangian is independent explicitly on time $t$ and azimuthal angle $\phi$, hence, one can obtain the two conserved constants:
\begin{equation}
\label{1-10}
p_{t}=\frac{\partial \mathcal{L} }{\partial \dot{t}}= H(r) f(r) \dot{t}= E,
\end{equation}
\begin{equation}
\label{1-11}
p_{\phi}=\frac{\partial \mathcal{L} }{\partial \dot{\phi}}= h(r) r^{2} \dot{\phi} sin^{2} \theta = L.
\end{equation}
The four-velocity of the time, the azimuthal angle, and the radial components can be obtained,
\begin{eqnarray}
\label{1-12}
&&\frac{{\rm d} t}{{\rm d} \tau}=\frac{1}{b}H(r)^{-1}f(r)^{-1},\\
\label{1-13}
&&\frac{{\rm d} r}{{\rm d} \tau}=\sqrt{\frac{1}{b^{2}H(r)^{2}} - \frac{1}{H(r)h(r)r^{2} }f(r)},\\
\label{1-14}
&&\frac{{\rm d} \phi}{{\rm d} \tau}=\pm\frac{1}{h(r) r^{2}},
\end{eqnarray}
where the symbol ``$\pm$" denotes that the counterclockwise ($-$) and clockwise ($+$) direction for the motion of photons. The $b$ is the impact parameter, satisfying $b = |L|/E$. The equations of motion for the null geodesic is
\begin{equation}
\label{1-15}
H(r) f(r)\dot{t}^{2}-{H(r)}{f(r)^{-1}}\Big(\frac{{\rm d} r}{{\rm d} \phi}\Big)^{2}\dot{\phi}^{2} - r^{2} h(r) \dot{\phi}^{2} = 0.
\end{equation}
Based on the method of Refs. \cite{Kruglov-2,31} and using Eqs. (\ref{1-12}) - (\ref{1-15}), the effective potential $V_{\rm e}$ of the EH BH can be written as
\begin{equation}
\label{1-16}
\Big(\frac{{\rm d} r}{{\rm d} \phi}\Big)^{2} = V_{\rm e} = r^{4}\Bigg(\frac{h(r)^{2}}{b^{2}H(r)^{2}}-\frac{f(r)h(r)}{H(r)r^{2}}\Bigg).
\end{equation}
Utilizing the effective potential function Eq. (\ref{1-16}) and the condition of unstable circular orbit $\frac{d V}{d r}=V=0$ \cite{Kruglov-2,31}, we have
\begin{eqnarray}
\label{1-17}
&&\frac{1}{b^{2}}=\frac{E^{2}}{L^{2}}=\frac{f(r)H(r)}{r^{2}h(r)},\\
\label{1-18}
&&r f(r) H(r) h'(r) + 2 f(r) H(r) h(r) \nonumber \\
&&- r f(r) h(r) H'(r) - r H(r) h(r) f'(r)=0,
\end{eqnarray}
According to above expression and Eqs. (\ref{1-4}) (\ref{1-7}) (\ref{1-8}), the numerical results of the EH BH event horizon radii, shadow radii and critical impact parameters for different parameter values are listed in Tab.\ref{Tab:1}. It is found that the increase of $Q$ value leads to the decrease of $r_{\rm +}$, $r_{\rm ph}$ and $b_{\rm ph}$, implying that the BH photon ring is shrunk inward the BH by increasing the magnetic charge. However, these key quantities are not sensitive to the EH parameter $a$.
\begin{table}[h]
\caption{The EH BH event horizon radius $r_{\rm +}$, shadow radius $r_{\rm ph}$, and critical impact parameter $b_{\rm ph}$ for different parameter values in case of the BH mass of $M=1$.}
\label{Tab:1}
\begin{center}
\setlength{\tabcolsep}{2.5mm}
\linespread{1cm}
\begin{tabular}[t]{|c|c|c|c|c|c|c|c|c|c|}
  \hline
  $Q$ & $0$ & $0.2$ & $0.4$ & $0.6$ & $0.8$ & $1.0$\\
  \hline
  $r_{\rm +}$   &  $2.00$   &  $1.98$  &  $1.92$  &  $1.80$  &  $1.60$   &   $1.12$  \\
   \hline
  $r_{\rm ph}$  &  $3.00$   &  $2.97$  &  $2.89$  &  $2.73$  &  $2.47$   &   $1.94$  \\
   \hline
  $b_{\rm ph}$  &  $5.20$ &  $5.16$  &  $5.05$  &  $4.85$  &  $4.53$   &   $3.94$  \\
  \hline
  \hline
  $a$ & $0$ & $0.2$ & $0.4$ & $0.6$ & $0.8$ & $1.0$\\
  \hline
  $r_{\rm +}$   &  $1.87$   &  $1.87$  &  $1.87$  &  $1.87$  &  $1.87$   &   $1.87$  \\
   \hline
  $r_{\rm ph}$  &  $2.82$   &  $2.82$  &  $2.82$  &  $2.82$  &  $2.82$   &   $2.82$  \\
   \hline
  $b_{\rm ph}$  &  $4.97$   &  $4.97$  &  $4.96$  &  $4.97$  &  $4.96$   &   $4.96$  \\
   \hline
\end{tabular}
\end{center}
\end{table}

\par
\begin{figure*}[htbp]
  \centering
  \includegraphics[width=5.2cm,height=5.2cm]{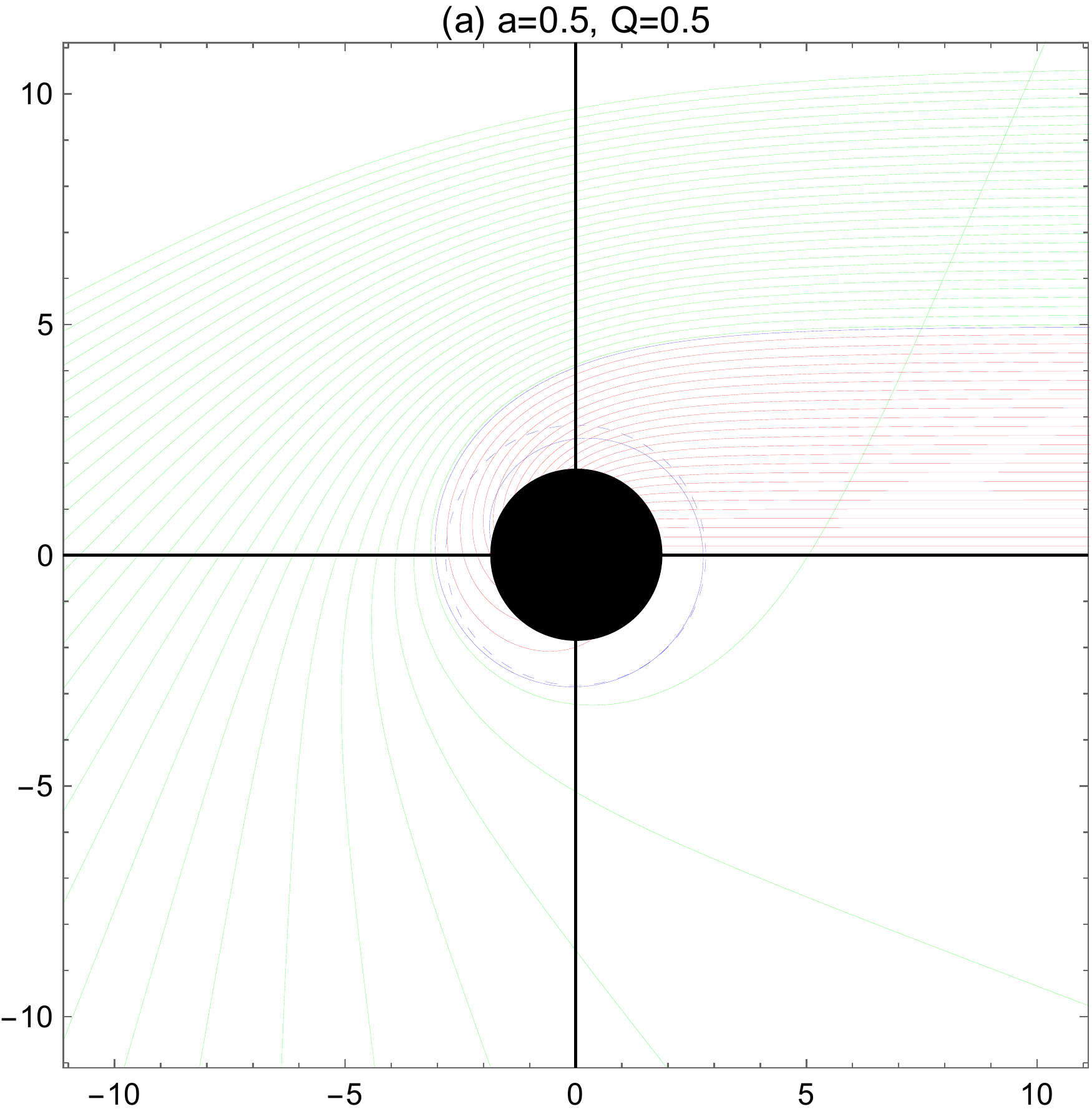}
  \hspace{0.5cm}
  \includegraphics[width=5.2cm,height=5.2cm]{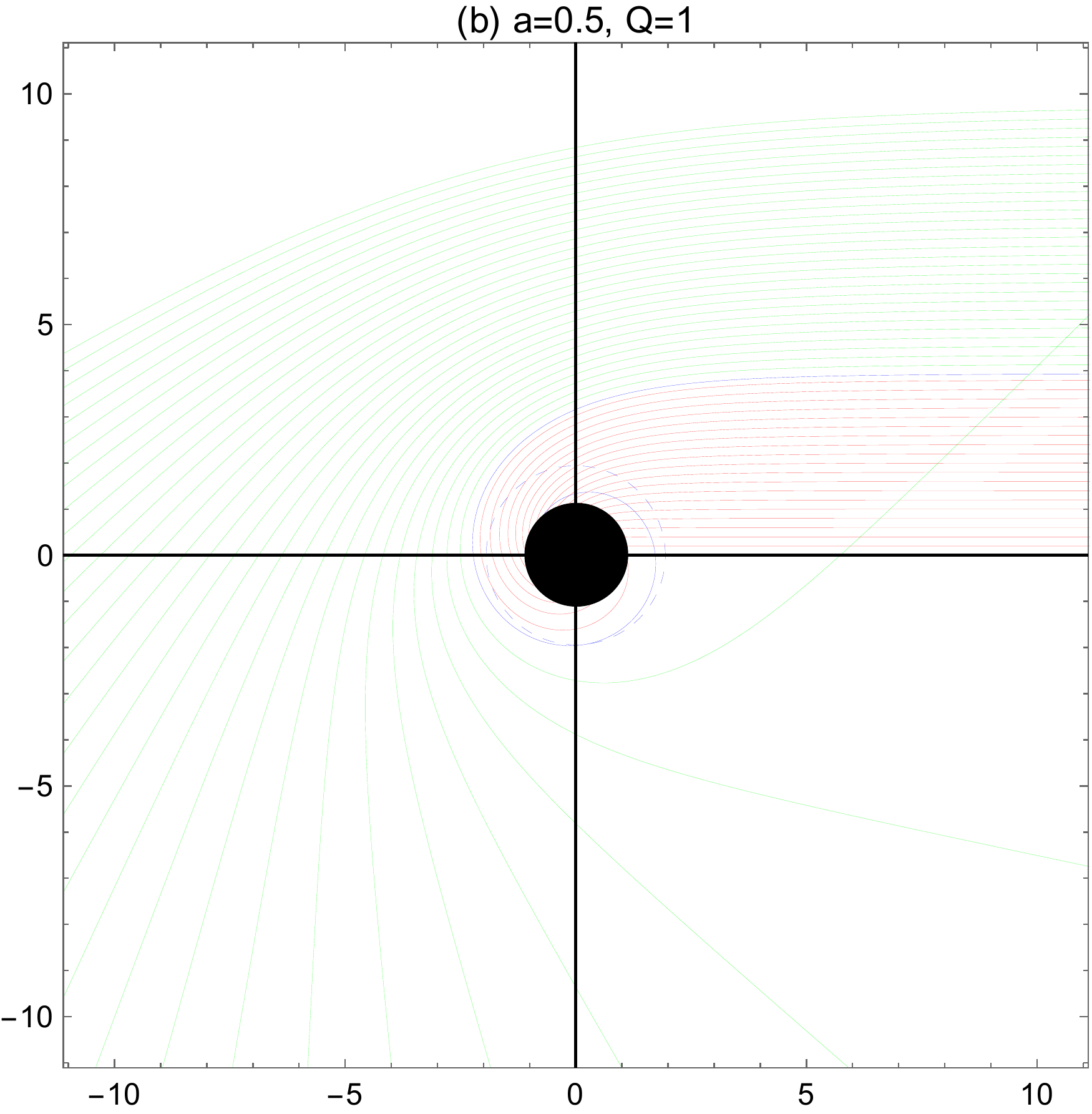}
  \hspace{0.5cm}
  \includegraphics[width=5.2cm,height=5.2cm]{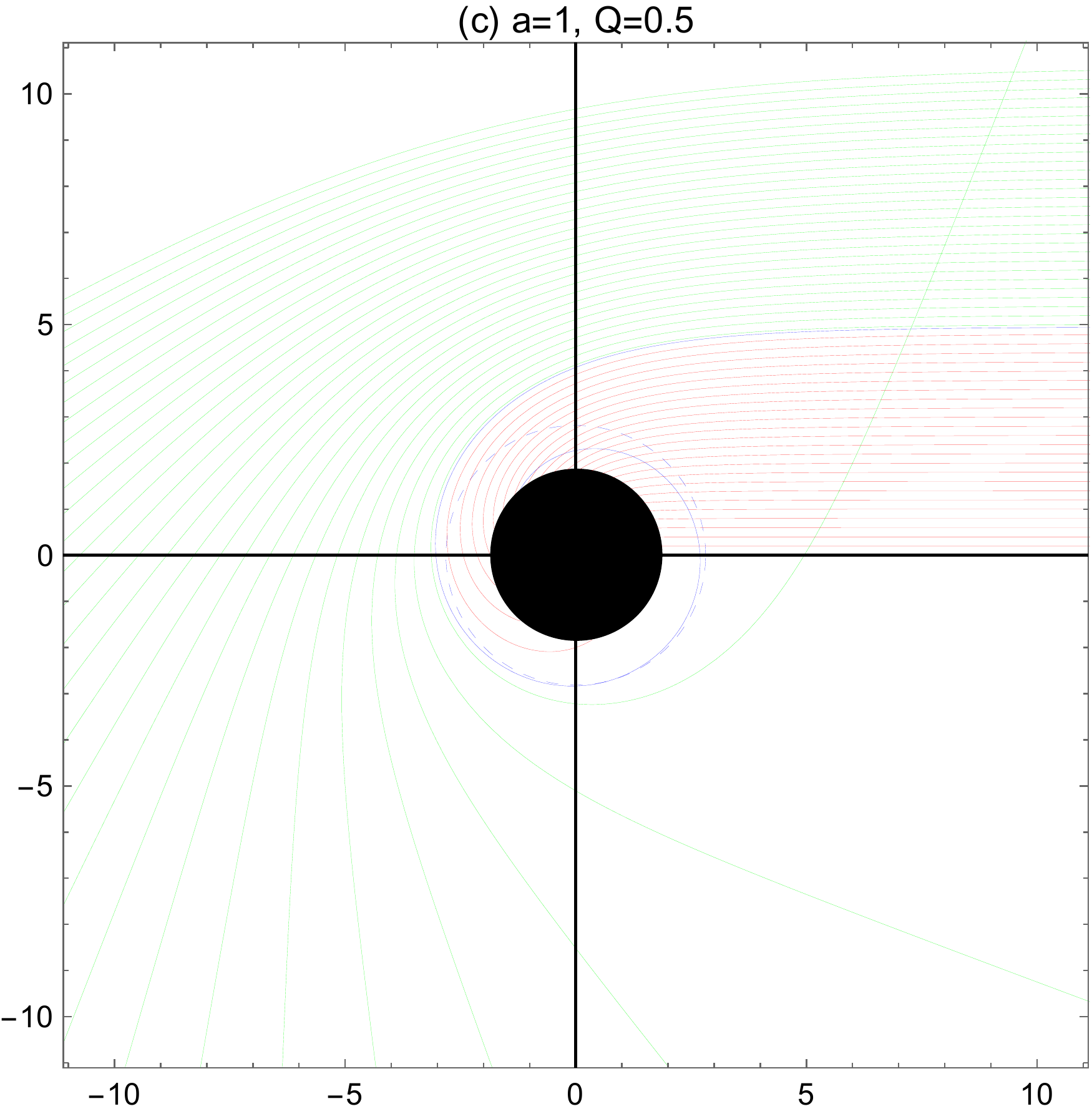}
  \caption {The trajectory of the light ray for different parameters in the polar coordinates $(r,\phi)$. The BH mass as $M=1$. {\em Panel (a)}-- magnetic charge $Q=0.5$ and EH parameter $a=0.5$, {\em Panel (b)}-- magnetic charge $Q=1$ and EH parameter $a=0.5$, and {\em Panel (c)}-- magnetic charge $Q=0.5$ and EH parameter $a=1$. The red lines, blue lines and green lines correspond to $b<b_{\rm ph}$, $b=b_{\rm ph}$ and $b>b_{\rm ph}$, respectively. The BH is shown as a black disk and photon orbit as a dashed blue line.}\label{fig:1}
\end{figure*}

Based on Eq. (\ref{1-16}), we can obtain that
\begin{equation}
\label{a-9}
\frac{{\rm d} r}{{\rm d} \phi} = \pm r^{2}\sqrt{\frac{h(r)^{2}}{b^{2}H(r)^{2}} - \frac{f(r)h(r)}{H(r)r^{2}}}.
\end{equation}
By introducing a parameter $u \equiv {1}/{r}$, the expression of $\Pi(u) \equiv \frac{{\rm d} u}{{\rm d} \phi}$ is obtained. Utilizing the ray-tracing code, the trajectory of the light ray for different parameter values are shown in Fig. \ref{fig:1}, showing that the radius of the black disk is smaller and the light rays are more gentle near the BH for a larger $Q$. The light density received by a distant observer increases with the increase of $Q$. Note that the deflection of the light ray is insensitive to the EH parameter.

\par
\begin{figure}[htbp]
  \centering
  \includegraphics[width=7.5cm,height=5cm]{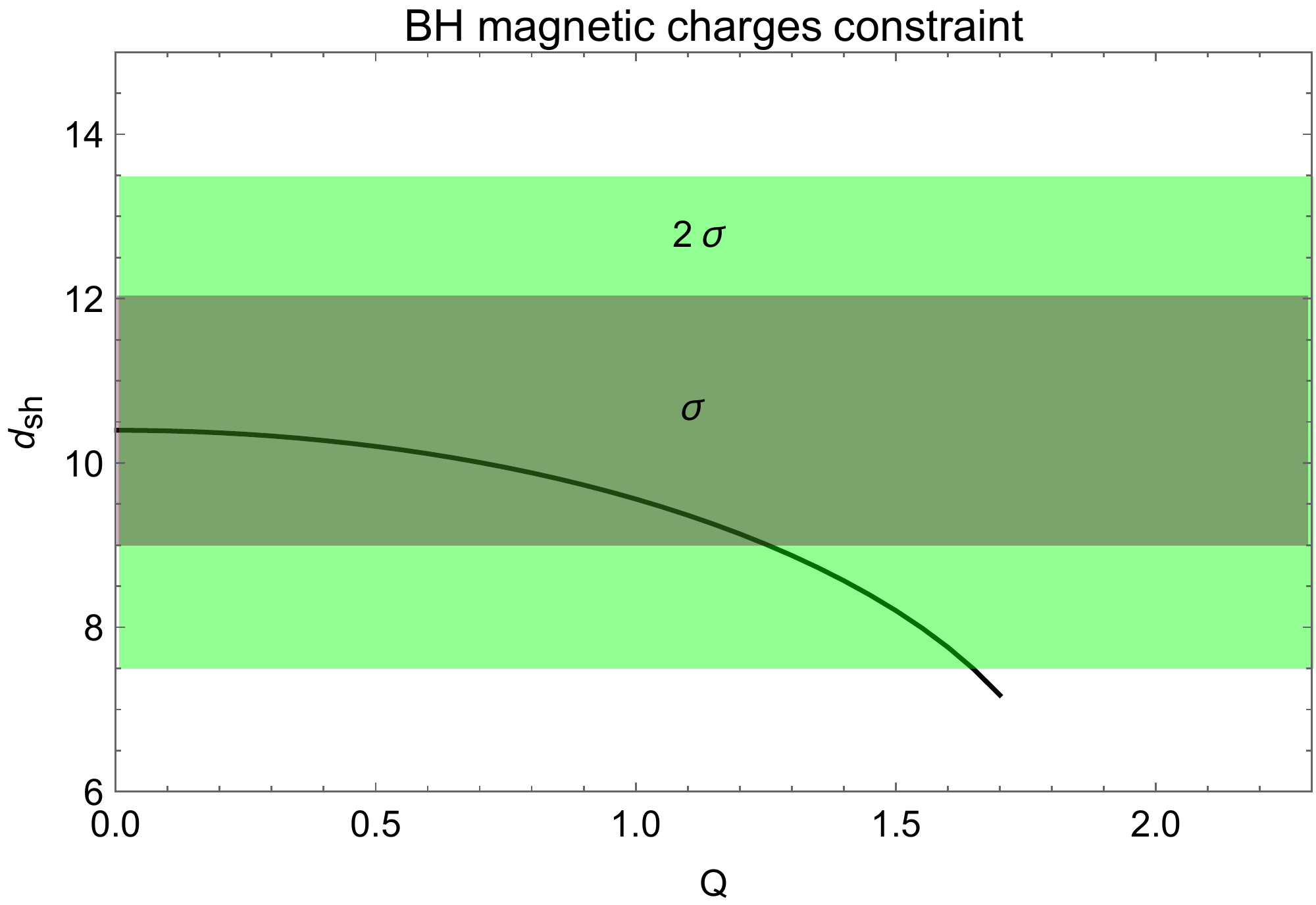}
  \caption {Shadow diameter of the EH BH as a function of magnetic charge. The diameter of the M87$^{*}$ estimated with the EHT observations and its uncertainties in $1\sigma$ ($2\sigma$) confidence levels are marked with a horizonal dashed-line and dark (or light) green shaded regions.}\label{fig:2}
\end{figure}

The magnetic charge as a free parameter, the BH shadow radius depends on it. Fig. \ref{fig:2} shows the BH shadow diameter $d_{\rm sh}$ as a function of the BH magnetic charge $Q$. One can observe that the $d_{\rm sh}$ decreases with increase of $Q$. When the $Q=0$, $d_{\rm sh}$ is $10.4 r_{\rm g}$, it is consistent with the Schwarzschild BH. Note that $d_{\rm sh}$ can be measured with the EHT data. The angular size of the shadow of M87$^{*}$ is $\delta = (42 \pm 3)$ $\mu$as, its distance is $D= 16.8_{-0.7}^{+0.8}$ Mpc, and its BH mass $M=(6.5 \pm 0.9) \times 10^{9}M_{\odot}$. The diameter of its shadow is given as $d_{\rm M87^{*}}\approx 11.0 \pm 1.5$ \cite{31,32}. Fig. \ref{fig:2} illustrates our result is consistent with that derived from the EHT observations within the observational uncertainty. Using the $1\sigma$ and $2\sigma$ confidence intervals of $d_{\rm M87^{*}}$, $g$ can be constrained as $Q \lesssim 1.28$ at $1 \sigma$ and $Q \lesssim 1.64$ at $2 \sigma$.

\section{\textbf{Optical appearance of the EH BH under the three accretion flow models}}
\label{sec:3}
\subsection{Static spherical accretion flow model}
\label{sec:3-1}
\par
When the BH is surrounded by a static, optically thin, and geometrically thin spherical accretion flow, the observed intensity ($\rm ergs^{-1}cm^{-2}str^{-1}Hz^{-1}$) with a frequency $\upsilon^{\rm s}_{\rm o}$ is given by \cite{Jaroszynski}
\begin{equation}
\label{3-1-1}
I^{\rm s}(b)=\int {g^{\rm s}}^{3} j(\upsilon^{\rm s}_{\rm e}) {\rm d}l_{\rm prop},
\end{equation}
where $g^{\rm s}\equiv \upsilon^{\rm s}_{\rm o}/\upsilon^{\rm s}_{\rm e}$ is the red-shift factor, $\upsilon^{\rm s}_{\rm e}$ is the intrinsic photon frequency, ${\rm d}l_{\rm prop}$ is the infinitesimal proper length, and $j(\upsilon^{\rm s}_{\rm e})$ is the emissivity per unit volume in the rest frame of the emitter. For the EH BH, the red-shift factor can be regarded as $g^{\rm s} \equiv H(r)^{1/2} f(r)^{1/2}$. Note that, we take into account the effective geometry induced by the EH non-linear electrodynamics here. We consider a simple case of the emission is monochromatic with rest-frame frequency $\upsilon_{\rm t}$. It emissivity has a radial profile as $1/r^{2}$, one can get
\begin{equation}
\label{3-1-2}
j(\upsilon^{\rm s}_{\rm e}) \propto \frac{\delta(\upsilon^{\rm s}_{\rm e}-\upsilon_{\rm t})}{r^2}.
\end{equation}
Under the EH BH context, the proper length measured in the rest frame of the emitter can be written as
\begin{eqnarray}
\label{3-1-3}
{\rm d}l_{\rm prop} &&= \sqrt{H(r)f(r)^{-1}{\rm d} r^{2} + h(r) r^{2} {\rm d} \phi^{2}}\nonumber\\
&&= \sqrt{\frac{H(r)}{f(r)} + h(r) r^{2} \Big(\frac{{\rm d} \phi}{{\rm d} r}\Big)^{2}}{\rm d} r.
\end{eqnarray}
Utilizing Eqs. (\ref{3-1-1})-(\ref{3-1-3}), the total observed intensity of the EH BH on the static spherical accretion background is obtained, i.e.
\begin{equation}
\label{3-1-4}
I^{\rm s}(b)=\int \frac{H(r)^{2}f(r)}{r}\sqrt{\frac{h(r)r^{2}}{h(r)-b^{2}f(r)H(r)}}{\rm d}r.
\end{equation}
\begin{figure*}[htbp]
  \centering
  \includegraphics[width=8.5cm,height=6cm]{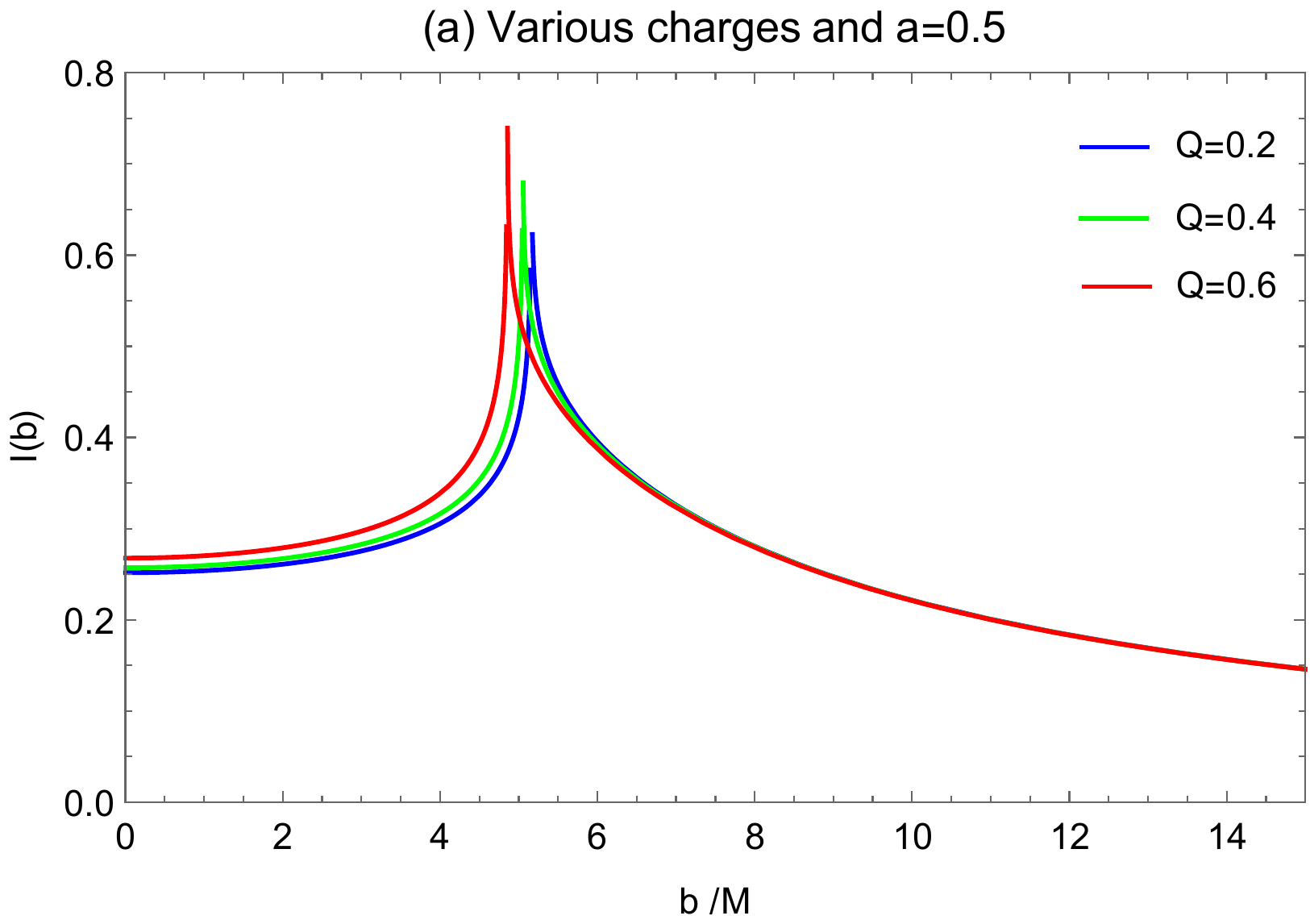}
  \hspace{0.2cm}
  \includegraphics[width=8.5cm,height=6cm]{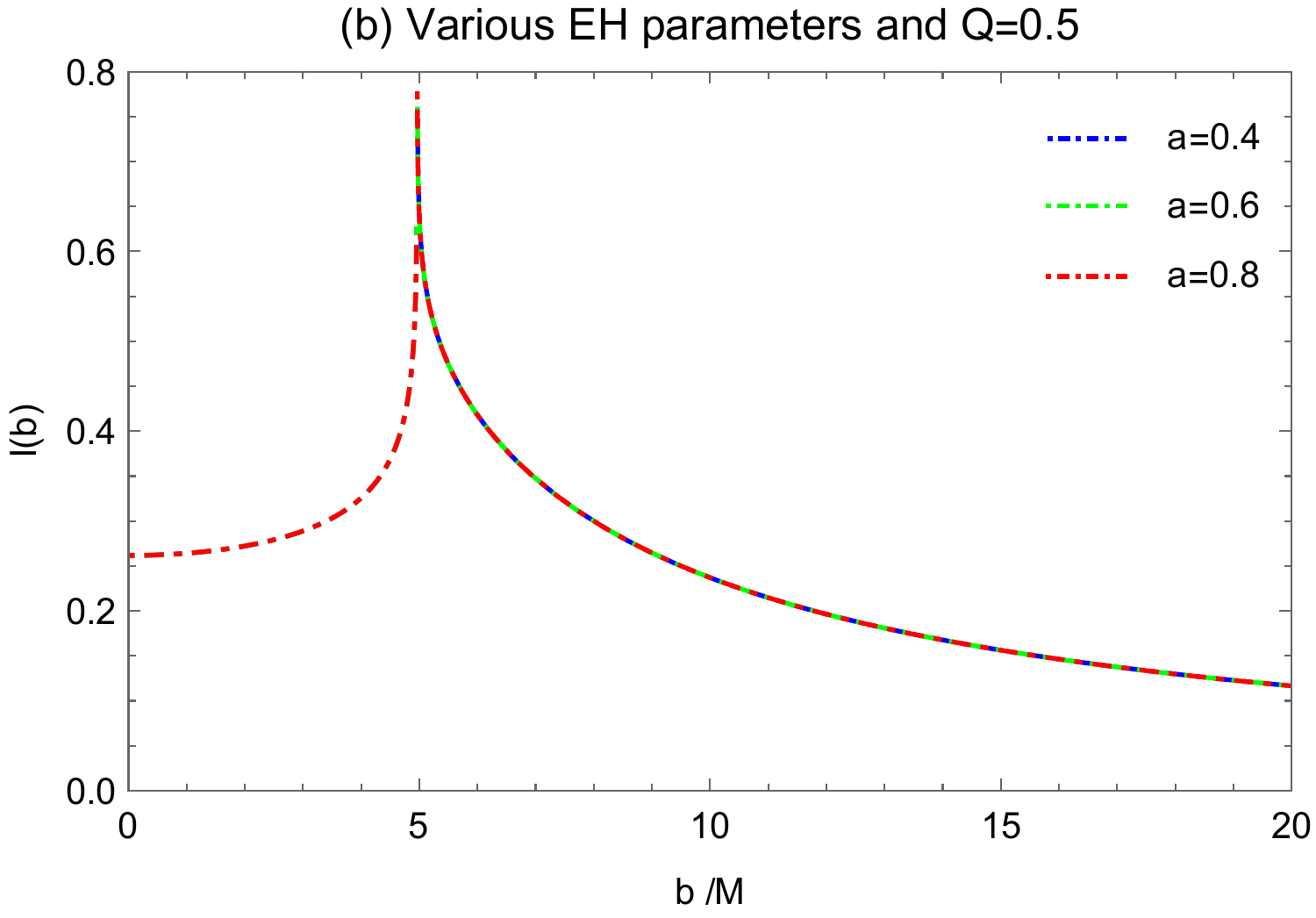}
  \caption {The total observed intensity $I^{\rm s}(b)$ as a function of impact parameter $b$ for the EH BH under a static spherical accretion flow context. {\em Panel (a)}-- different BH magnetic charges and {\em Panel (b)}-- different EH parameters. The BH mass is $M=1$.}\label{fig:3}
\end{figure*}
\begin{figure*}[htbp]
  \centering
  \includegraphics[width=5.2cm,height=5.2cm]{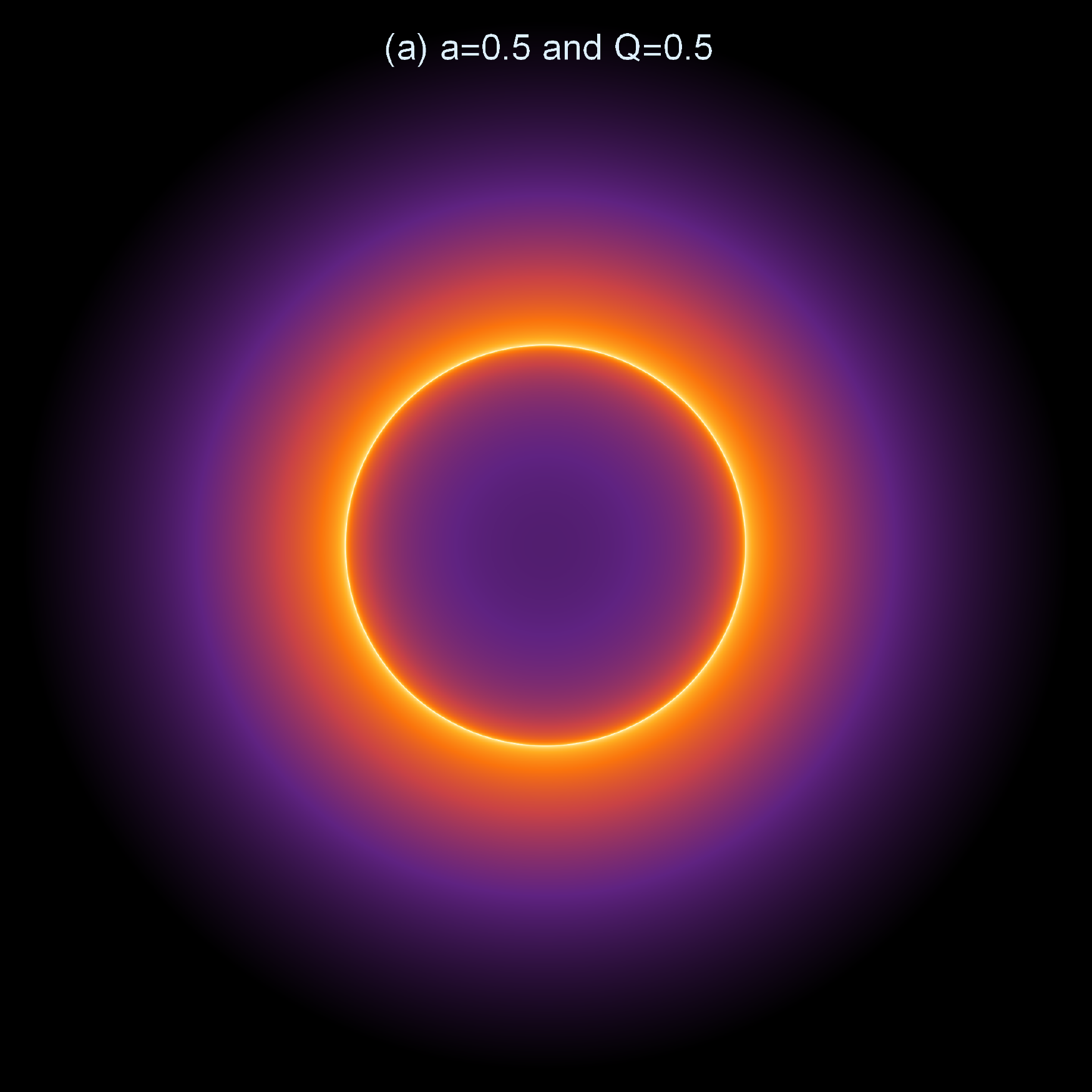}
  \hspace{0.5cm}
  \includegraphics[width=5.2cm,height=5.2cm]{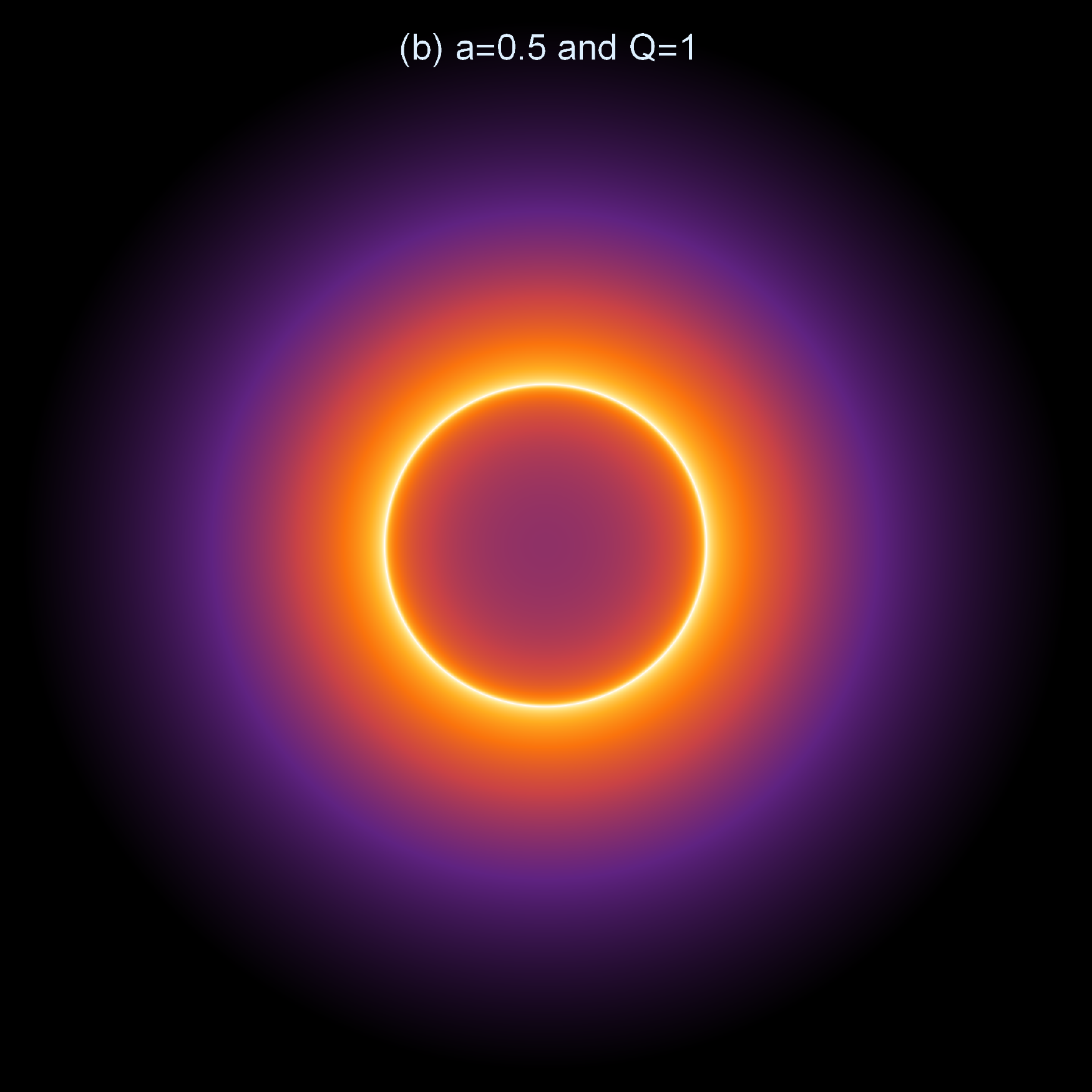}
  \hspace{0.5cm}
  \includegraphics[width=5.2cm,height=5.2cm]{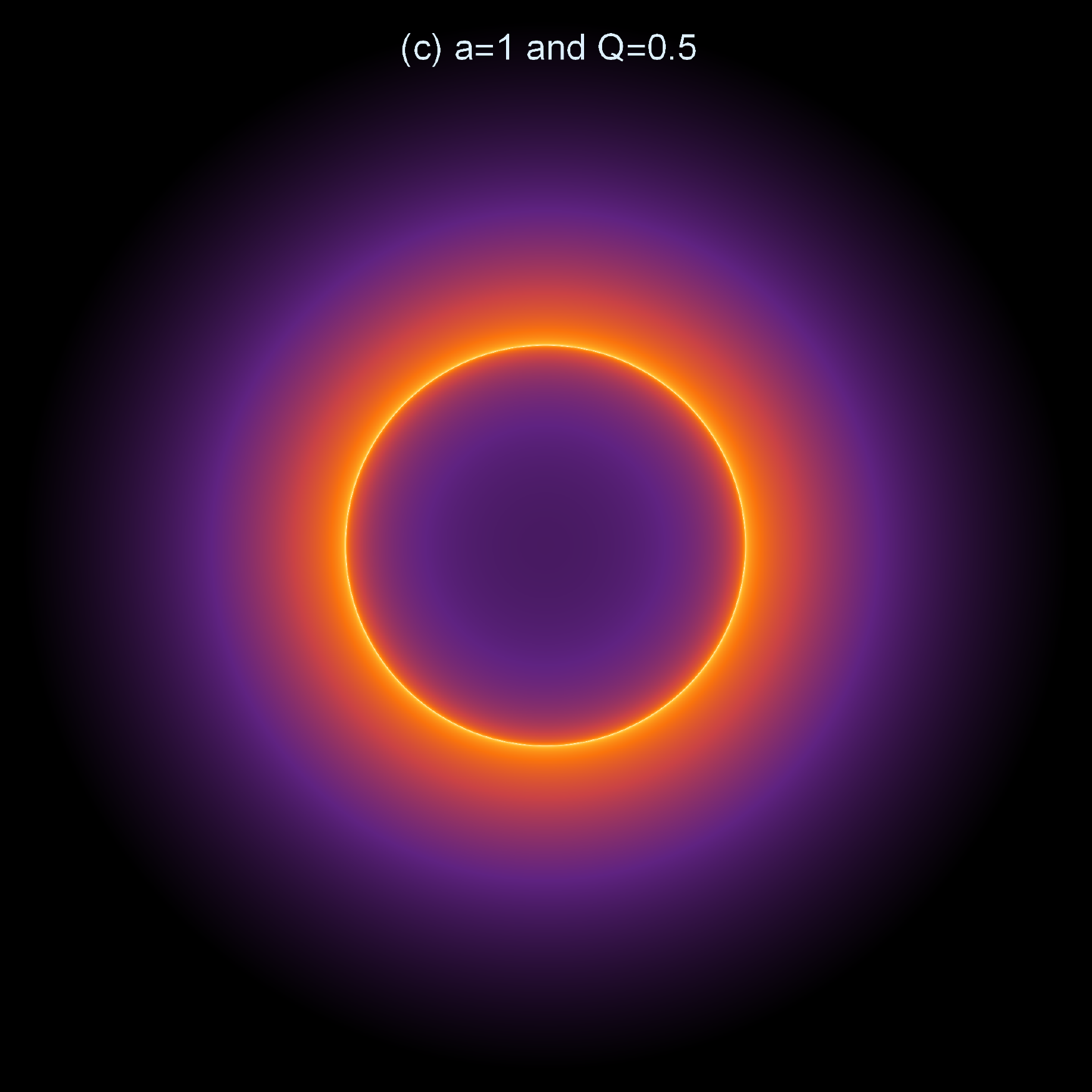}
  \caption {Two-dimensional images of shadows and photon rings of the EH BH with a static spherical accretion flow. {\em Panel (a)}-- magnetic charge $Q=0.5$ and EH parameter $a=0.5$, {\em Panel (b)}-- magnetic charge $Q=1$ and EH parameter $a=0.5$, and {\em Panel (c)}-- magnetic charge $Q=0.5$ and EH parameter $a=1$.}\label{fig:4}
\end{figure*}

\par
Fig. \ref{fig:3} shows that the observed intensity $I^{\rm s}(b)$ as a function of the impact parameter $b$ for several representative parameter values. The observed intensity first ascended with the impact parameter, and reached the peak at $b_{\rm ph}$, which corresponds to the photon ring of the BH. For optically thin sources, the path length of a ray through the source is a proxy for the intensity picked up by that ray. Due to the light ray rotating around BH many times in the BH photon ring orbit, the optical path is infinite. Thus, a distant observer sees the strongest luminosity at the critical impact parameter. The peak value of the observed intensity increase with an increase of the BH magnetic charge when the EH parameter is a constant, and the corresponding $b_{\rm ph}$ get smaller. The EH parameter does not affect the luminosity change, and the intensity curve is always constant whether $a$ increases or decreases.

\par
Fig. \ref{fig:4} expounds that the two-dimensional shadows cast by this BH in the celestial coordinates. One can observe that a bright ring surrounds a central dark area. The shadow is not a totally dark region with zero intensity since the perspective of the shadows seen by a distant observer is occupied by the photons radiated from the bright accretion flow. The luminosities of the smaller magnetic charge BH shadows and photon rings are darker than that of the larger magnetic charge BH since the BH magnetic charge makes the BH photon ring can capture more photons from the accretion flow. Thus, the EH shadows and photon rings' luminosities increase gradually as the magnetic charge increases. Consistent with the results in Tab. \ref{Tab:1}, the change of $a$ does not affect the size and luminosity of the shadow, which indicates that the luminosity of the EH BH is independent of the one-loop corrections to QED effect.

\subsection{Infalling spherical accretion flow model}
\label{sec:3-2}
\par
We consider the EH BH is surrounded by a radial infalling and optically/geometrically thin spherical accretion flow. In this case, the equation of the observed intensity on the static spherical accretion flow is still valid (Eq. \ref{3-1-1}). However, the redshift factor is different from the above situation, it can be written as \cite{Bambi}
\begin{equation}
\label{3-2-1}
g^{\rm i}=\frac{k_{\rm \alpha}u_{\rm o}^{\rm \alpha \rm i}}{k_{\rm \beta}u_{\rm e}^{\rm \beta \rm i}},
\end{equation}
in which $k_{\rm \mu}$ is the four-velocity of the photons, $u_{\rm o}^{\rm \mu \rm i}$ is the four-velocity of an observer, and $u_{\rm e}^{\rm \mu \rm i}$ is the four-velocity of the accretion flow. According to Eqs. (\ref{1-12})-(\ref{1-14}), one can obtain the $k_{\rm t}$ is a constant, that is, $k_{\rm t} \equiv {1}/{b}$. The $k_{\rm r}$ comes from $k_{\rm\beta}k^{\rm\beta}$, we have
\begin{equation}
\label{3-2-2}
\frac{k_{\rm r}}{k_{\rm t}}=\pm \frac{1}{H(r)f(r)}\sqrt{1-\frac{b^{2}H(r)}{h(r)r^{2}}},
\end{equation}
where the symbol ``$\rm \pm$" corresponds the photon is approaching ($\rm +$) or away ($\rm -$) from the BH. The four-velocity of the accretion flow as
\begin{eqnarray}
\label{3-2-3}
&&u_{\rm e}^{\rm t \rm i}=\frac{1}{H(r) f(r)},~~u_{\rm e}^{\rm \theta \rm i}=u_{\rm e}^{\rm \varphi \rm i}=0,\\
\label{3-2-4}
&&u_{\rm e}^{\rm r \rm i}=-\sqrt{1-H(r) f(r)}.
\end{eqnarray}
The redshift factor $g^{\rm i}$ in this scenario can be written as
\begin{equation}
\label{3-2-5}
g^{\rm i}=\frac{1}{u_{\rm e}^{\rm t \rm i}+\Big(\frac{k_{\rm r}}{k^{\rm i}_{\rm e}}\Big)u_{\rm e}^{\rm r \rm i}}.
\end{equation}
The infinitesimal proper length is
\begin{equation}
\label{3-2-6}
{\rm d}l_{\rm prop}=k_{\rm \beta}u_{\rm e}^{\rm \beta \rm i} {\rm d} \tau = \frac{k_{\rm t}}{g^{\rm i}|k_{\rm r}|}{\rm d}r.
\end{equation}
Hence, one can obtain that the total observed intensity of the EH BH under the infalling spherical accretion flow context:
\begin{equation}
\label{3-2-7}
I^{\rm i}(b) = \int \frac{{g^{\rm i}}^{2}}{H(r) f(r) r^{2}} \sqrt{1-\frac{b^{2} H(r)}{h(r) r^{2}}} {\rm d} r.
\end{equation}
\begin{figure*}[htbp]
  \centering
  \includegraphics[width=8.5cm,height=6cm]{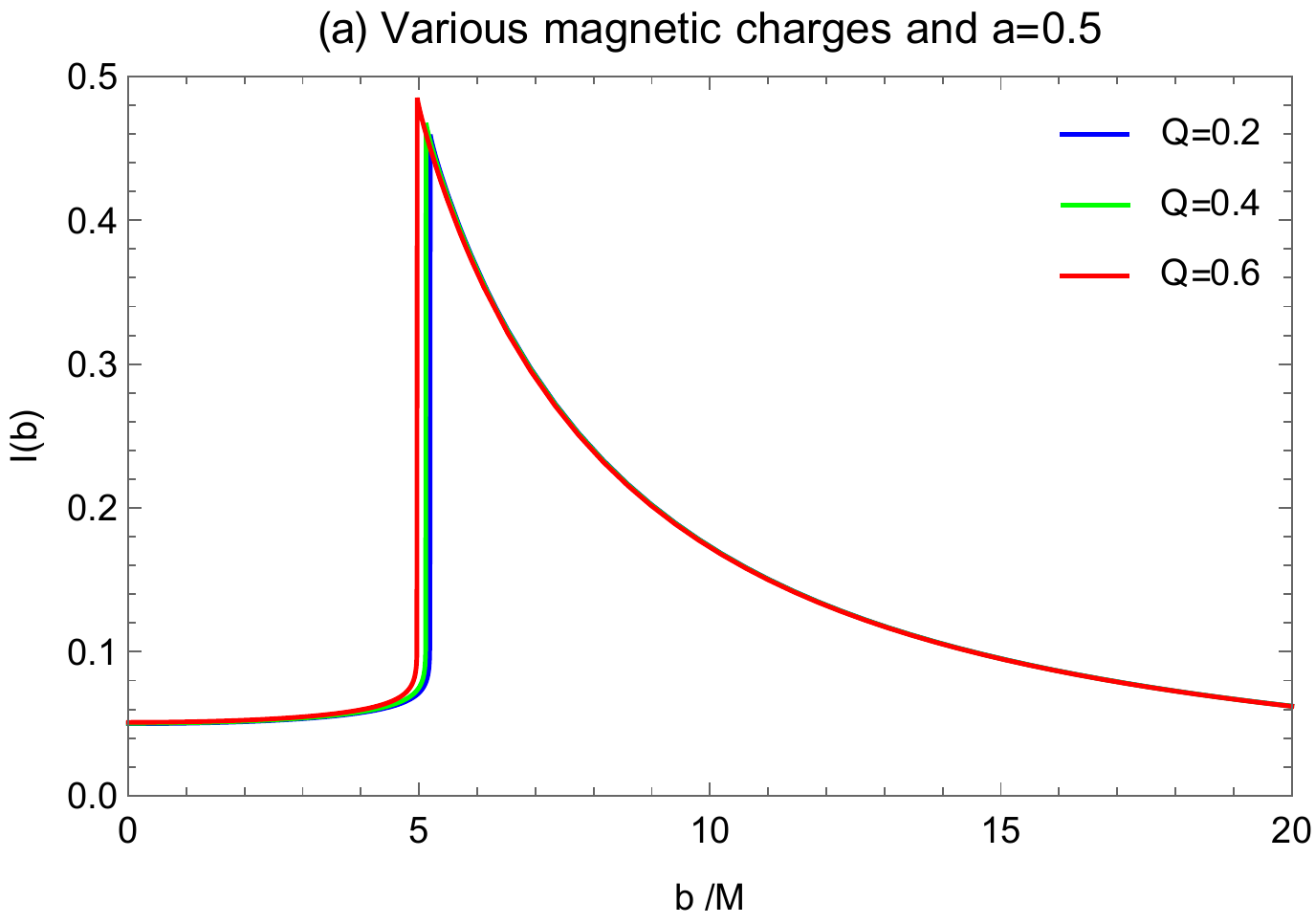}
  \hspace{0.2cm}
  \includegraphics[width=8.5cm,height=6cm]{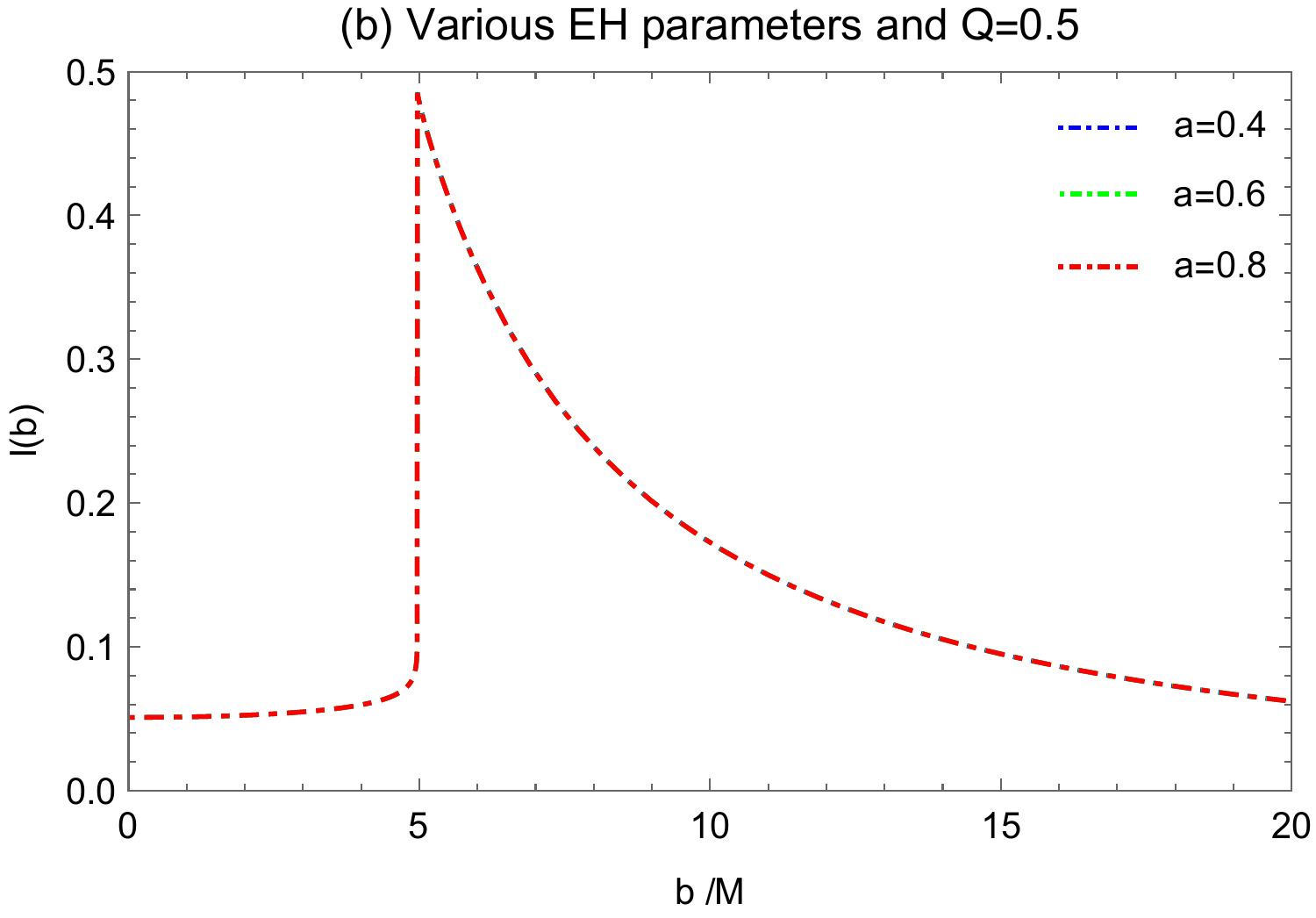}
  \caption {The total observed intensity $I^{\rm i}(b)$ as a function of impact parameter $b$ for the EH BH under an infalling spherical accretion flow context. {\em Panel (a)}-- different BH magnetic charges and {\em Panel (b)}-- different EH parameters. The BH mass is $M=1$.}\label{fig:5}
\end{figure*}
\begin{figure*}[htbp]
  \centering
  \includegraphics[width=5.2cm,height=5.2cm]{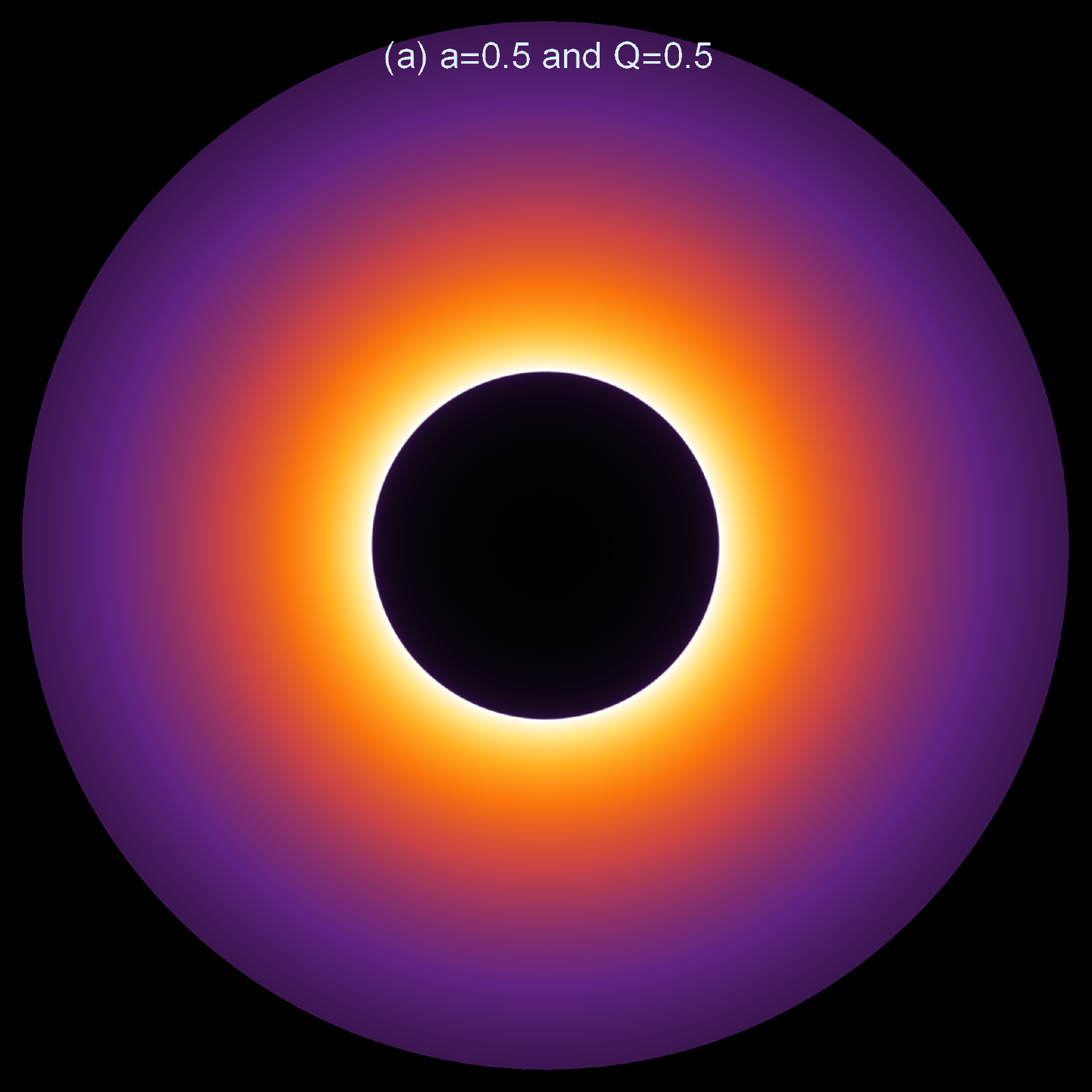}
  \hspace{0.5cm}
  \includegraphics[width=5.2cm,height=5.2cm]{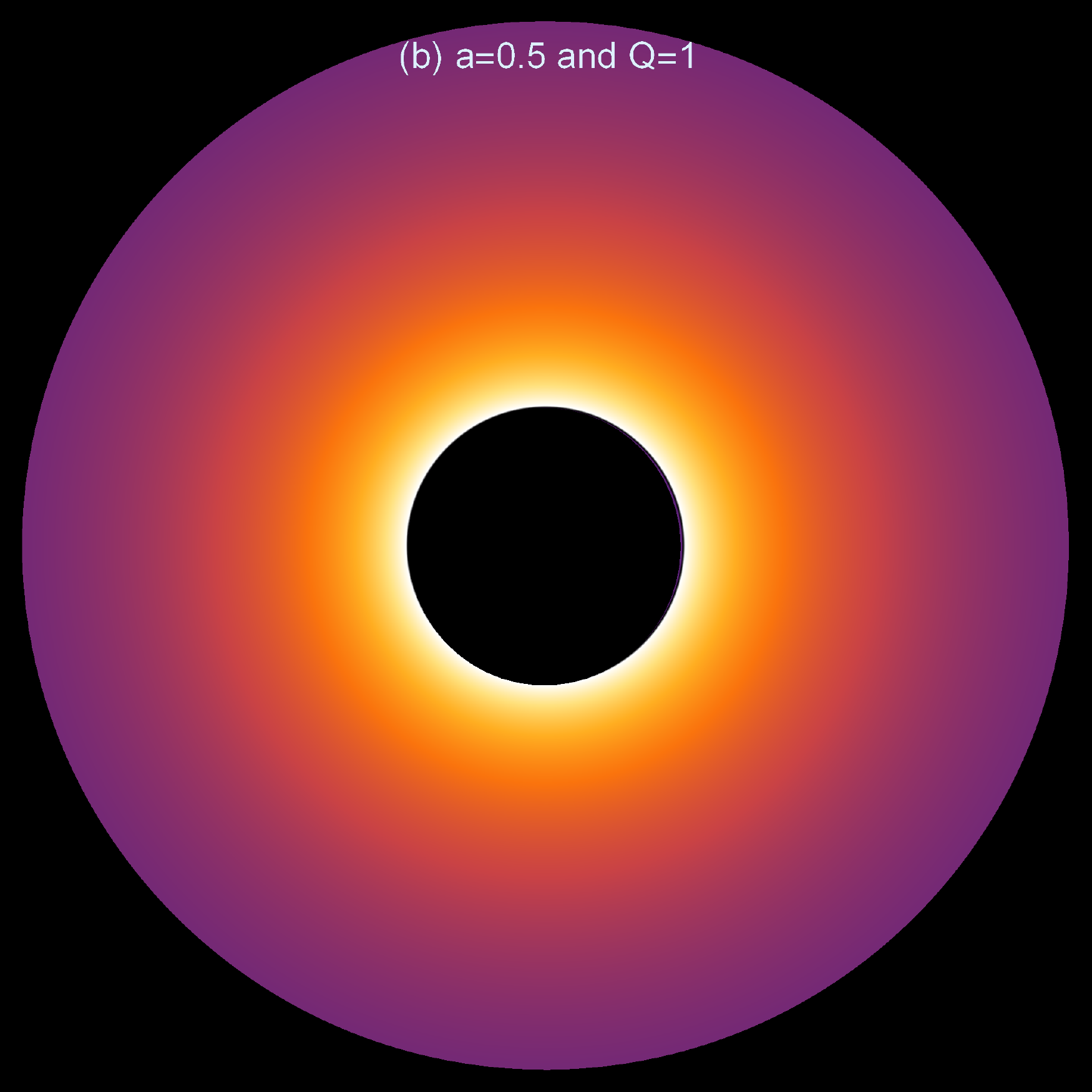}
  \hspace{0.5cm}
  \includegraphics[width=5.2cm,height=5.2cm]{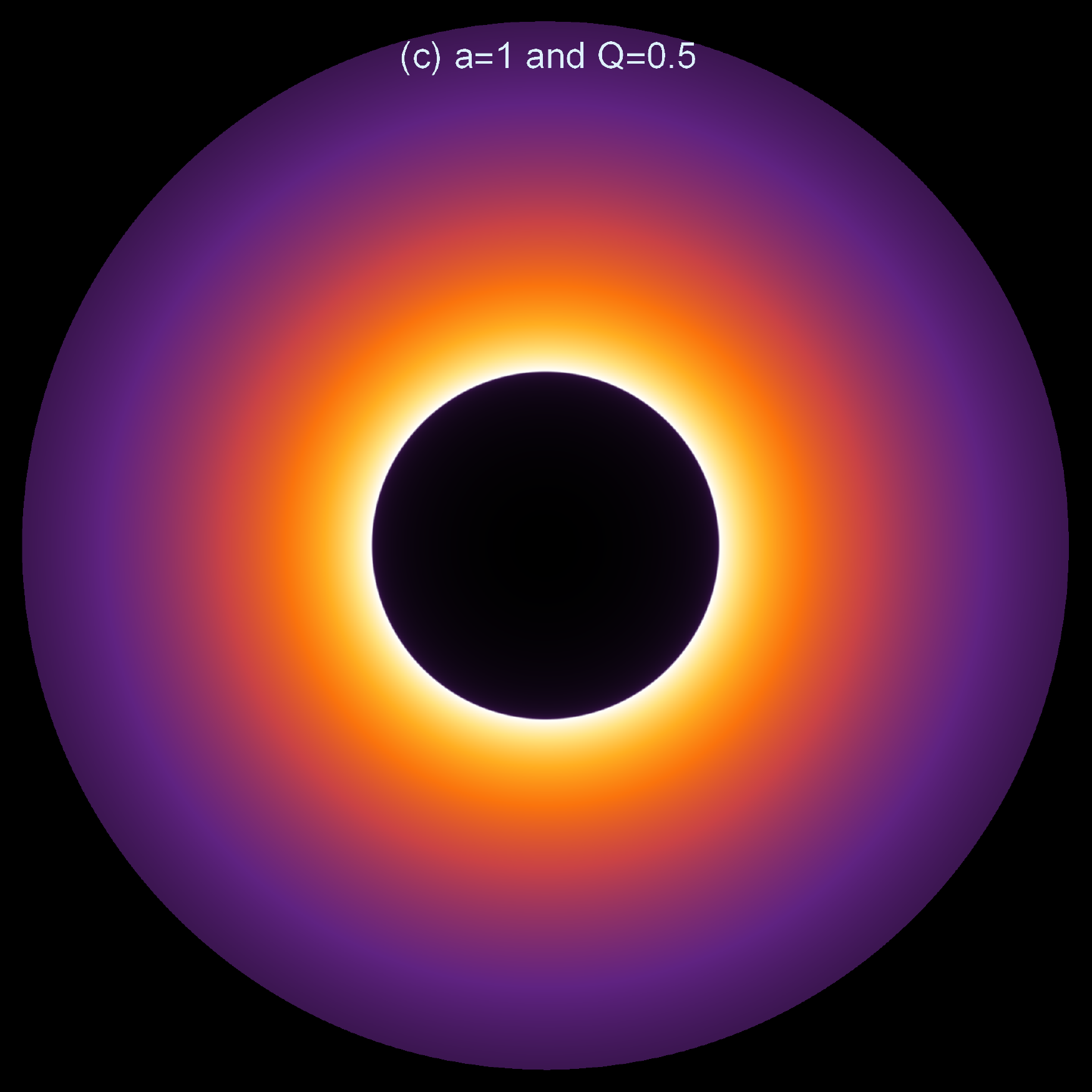}
  \caption {Two-dimensional images of shadows and photon rings of the EH BH with an infalling spherical accretion flow. {\em Panel (a)}-- magnetic charge $Q=0.5$ and EH parameter $a=0.5$, {\em Panel (b)}-- magnetic charge $Q=1$ and EH parameter $a=0.5$, and {\em Panel (c)}-- magnetic charge $Q=0.5$ and EH parameter $a=1$.}\label{fig:6}
\end{figure*}

\par
Fig. \ref{fig:5} shows that the total observed intensity function, the function curves shows similar feature to Fig. \ref{fig:3}, but the observed intensity has an extremely sharp rise before the peak. Fig. \ref{fig:6} illustrates that the two-dimensional shadows cast on the celestial coordinates in this scenario. Our result indicates that the size and position of the EH BH shadows do not change in both of these accretion flows.

\vskip 0.2cm
~~~~~~~~~~~~~~~~~~~~~~~~~~~~~~~~\emph{\textbf{Summary}}:
\vskip 0.2cm
The accretion flow property is crucial to the optical appearance of the BH. For the static and infalling spherical emitters, the luminosities of the photon rings with different magnetic charges are listed in Tab. \ref{Tab:2}. One can see that the total observed intensity in the static spherical accretion flow scenario leads than that of the infalling spherical accretion flow under same parameters.
\begin{table}[h]
\caption{The total observed intensity of the EH BH with static and infalling spherical accretion flows under different parameter values for $M=1$.}
\label{Tab:2}
\begin{center}
\setlength{\tabcolsep}{1.5 mm}
\linespread{0.1cm}
\begin{tabular}[t]{|c|c|c|c|c|c|c|c|c|c|}
  \hline
  $Q$     &     $0$    &    $0.2$    &    $0.4$     &     $0.6$     &     $0.8$     &     $1.0$  \\
  \hline
  $Static$ &  $0.689$   &  $0.704$    &  $0.732$  &  $0.791$   &   $0.812$  &   $0.913$  \\
   \hline
  $Infalling$  &  $0.457$   &  $0.461$    &  $0.475$  &  $0.500$   &   $0.538$  &   $0.635$\\
   \hline
\end{tabular}
\end{center}
\end{table}

\par
Compared with the static and infalling spherical emitters, we suggest that: \textit{i)} the size and position of the BH shadow does not change in case of static and infalling spherical accretion flows, implying that the BH shadow is a signature of space-time geometry; \textit{ii)} the BH shadow with infalling spherical accretion flow is darker than that of the static spherical accretion flow in the central region since the doppler-effect caused by the radial infalling accretion flow; \textit{iii)} the influence of the magnetic charge $Q$ on luminosity is basically the same under the two spherical accretion flows, i.e. the peak value of intensity increase with an increase of the BH magnetic charge when the EH parameter is a constant; \textit{iv)} QED effect affects the optical appearance of BH. Its existence makes the optical morphology of the EH BHs completely different from the RN BHs.

\par
\begin{figure*}[htbp]
  \centering
  \includegraphics[width=5.2cm,height=5.2cm]{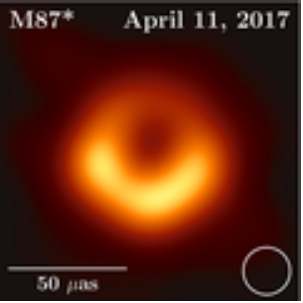}
  \hspace{0.5cm}
  \includegraphics[width=5.2cm,height=5.2cm]{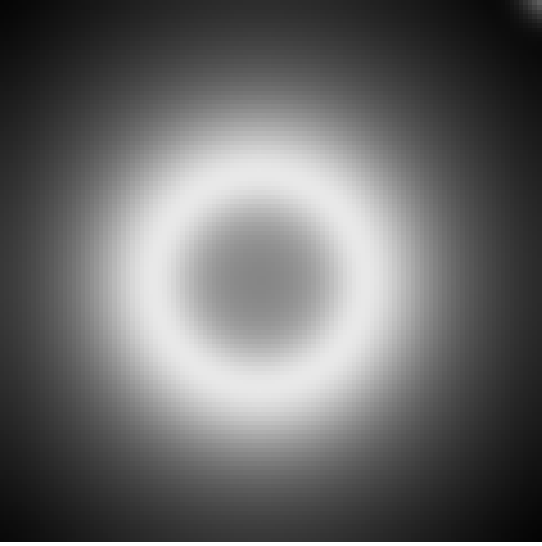}
  \hspace{0.5cm}
  \includegraphics[width=5.2cm,height=5.2cm]{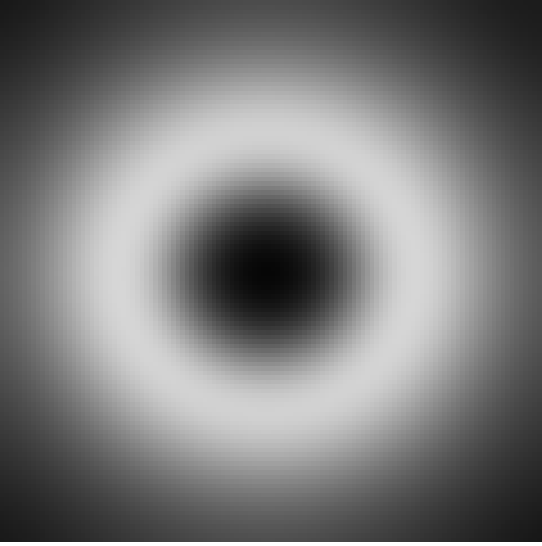}
  \caption {Blurred two-dimensional images utilizing a Gaussian filter with a standard deviation of 1/12 the field of view for the two spherical accretion flows.}\label{fig:7}
\end{figure*}

Additionally, we blur the two-dimensional image and correspond roughly to the EHT resolution. This simple blur does not correspond to the EHT image reconstruction and can only offer a rough illustration of the EHT resolution. From Fig. \ref{fig:7}, one can observe that although the observed shadow luminosity is different, the size of the BH shadow does not change for both static and infalling spherical accretion flow. These results suggest that the BH shadow size depends on the geometric space-time and the luminosity of the BH shadow relies on the accretion flow models. Compared with the results of the EHT, the shadow of the EH BH presents a complete circle, and the photon rings are evenly distributed around the shadow. According to Refs. \cite{2,3,4,5,6,7,8,9}, the M87$^{*}$ image shows that a supermassive rotating BH space is illuminated by a magnetically arrested accretion disk. This rotation results in the accumulation of brightness in the southwest of the EHT image. In the follow-up works, we will further describe the optical appearance of the rotating BH solutions within the framework of the EH theory. Inspired by these results, we investigate the optical appearance of the EH BH surrounded by an optically and geometrically thin disk accretion flow in the next section.

\subsection{Thin disk accretion flow model}
\label{sec:3-3}
\par
Assuming that the BH is illuminated by an accretion disk in the equatorial plane, the disk emits isotropically in the rest frame of static worldlines, and the observer is at the north pole. According to the definition of the total number of light orbits, the light trajectories emitted from the north pole direction can be subdivided into direct emission, lensing ring, and photon ring \cite{14}. When the light trajectories intersect the accretion disk just once and finally fall on the front, the corresponding light ray forms the direct emission. When the light trajectories intersect the accretion disk twice (the light breaks through thin disk) and finally fall on the back of the disk, which is the lensing ring. In this path, the light picks up additional brightness from the second intersection between the light trajectories and the accretion disk. The light arrives at the front side of the thin accretion disk once again for the photon ring, leading to additional brightness from the three intersections between the light and the disk. Hence, the total observed intensity should be the sum of those intensities.

\par
Based on the Liouville's theorem, $I_{\rm em}/({\upsilon_{\rm em}})^{3}$ is conserved in the direction of light propagation, where $I_{\rm em}$ and $\upsilon_{\rm em}$ delegate the radiation intensity and frequency of the accretion disk, respectively. An observer in infinity receive the specific intensity $I^{\rm d}_{\rm o}$ with red-shifted frequency $\upsilon^{\rm d}_{\rm o} \equiv \sqrt{f} \upsilon_{\rm em}$. Based on these results, we have
\begin{equation}
\label{3-3-1}
\frac{I^{\rm d}_{\rm o}}{(\upsilon^{\rm d}_{\rm o})^{3}}=\frac{I_{\rm em}}{(\upsilon_{\rm em})^{3}}.
\end{equation}
For the EH BH under the thin disk accretion flow context, the observed intensity for a specific frequency can be given by
\begin{eqnarray}
\label{3-3-2}
I^{\rm d}_{\rm o}(r)= H(r)^{{3}/{2}} f(r)^{{3}/{2}}I_{\rm em}(r).
\end{eqnarray}
The total intensity is an integral over all frequencies:
\begin{eqnarray}
\label{3-3-3}
I_{\rm S}(r)&&=\int I^{\rm d}_{\rm o}(r) {\rm d} \upsilon^{\rm d}_{\rm o}\nonumber \\
&&= \int H(r)^{2} f(r)^{2} I_{\rm em}(r) {\rm d} \upsilon_{\rm em}= H(r)^{2} f(r)^{2} I_{\rm e}(r),
\end{eqnarray}
where $I_{\rm e}(r) \equiv\int I_{\rm em}(r) {\rm d} \upsilon_{\rm em}$ is the total radiation intensity of the thin disk accretion flow. Hence, the total observed intensity of the EH BH in this situation can be written as
\begin{eqnarray}
\label{3-3-4}
I^{\rm d}(b)= \sum\limits_{n} H(r)^{2} f(r)^{2} I_{\rm e}|_{r=r_{\rm n}(b)},
\end{eqnarray}
where $r_{\rm n}(b)$  is the so-called transfer function which represents the radial coordinate of the $n_{\rm th}$ intersection between the light with impact parameter $b$ and the accretion disk. The slope of the $r_{\rm n}(b)$ - ${\rm d}r/{\rm d}b$ - is defined as the (de)magnification factor.

\par
\begin{figure*}[htbp]
  \centering
  \includegraphics[width=8.5cm,height=6cm]{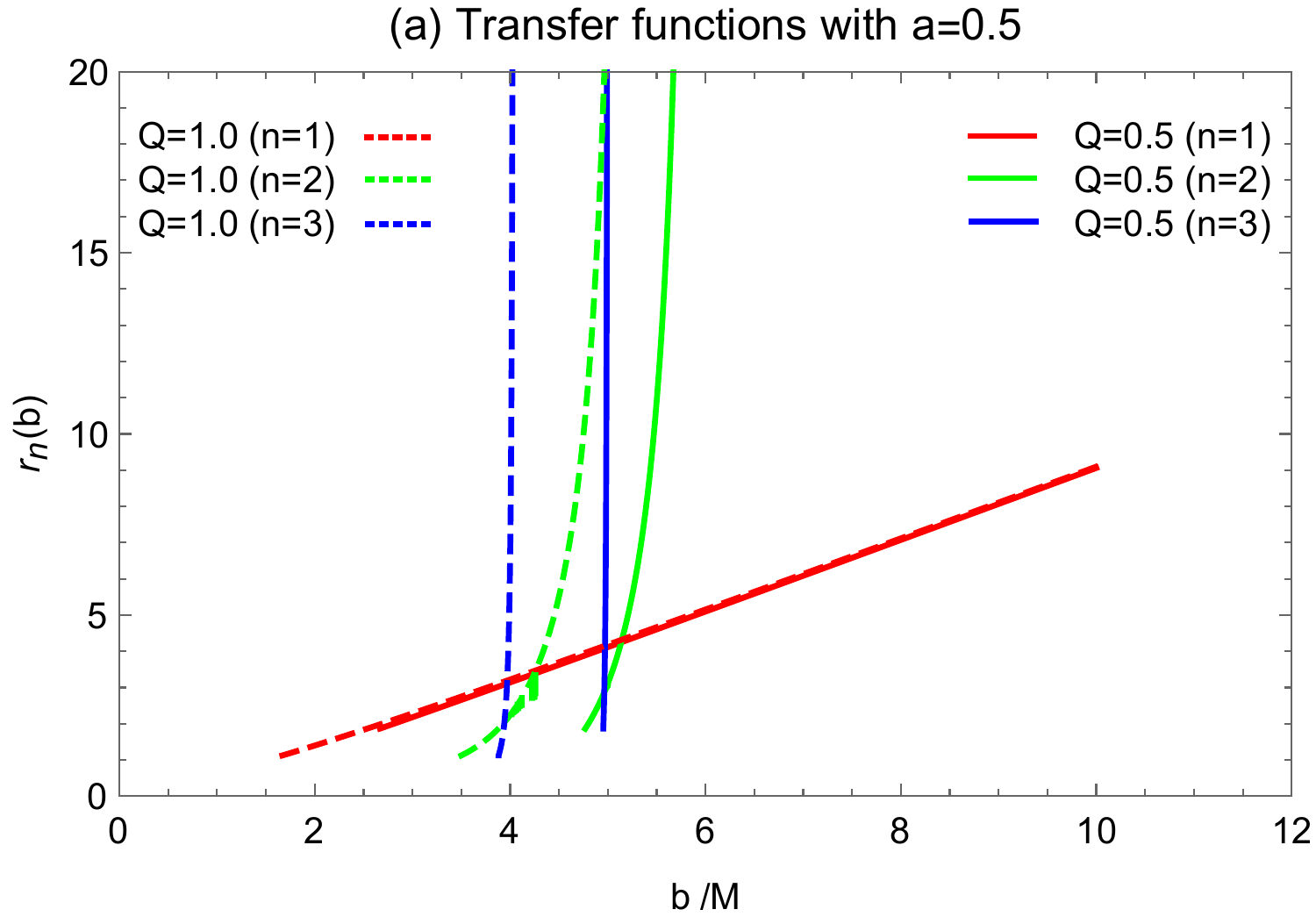}
  \hspace{0.2cm}
  \includegraphics[width=8.5cm,height=6cm]{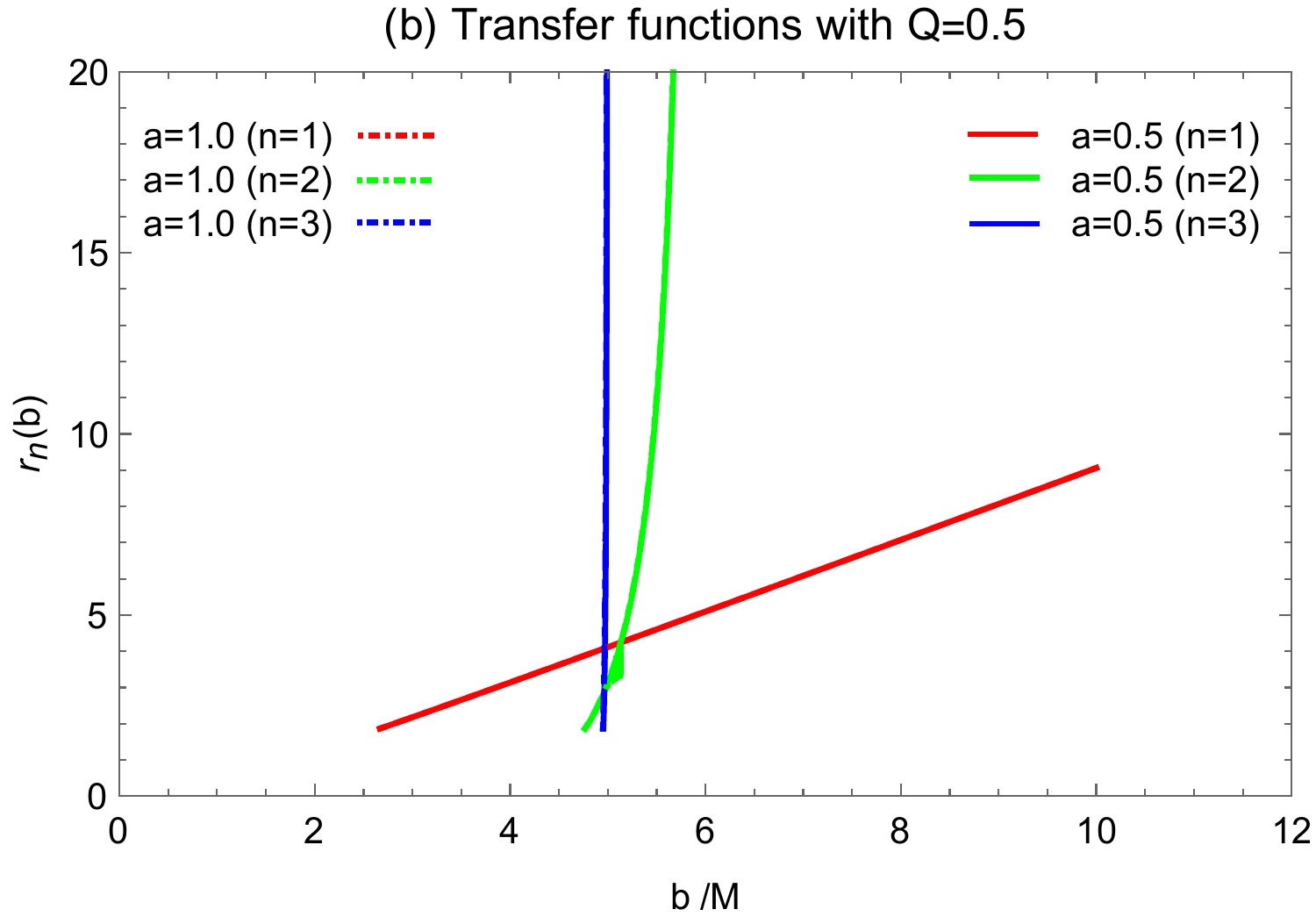}
  \caption {The $r_{\rm n}(b)$ as a function $b$ for different $Q$ and $a$. The BH mass is taken as $M=1$.}\label{fig:8}
\end{figure*}

Fig. \ref{fig:8} shows that $r_{\rm n}(b)$ as a function of $b$ for different parameter values. For the case of $n=1$, the $r_{\rm n}(b)$ is a linear function with a slope of 1, indicating that $r_{\rm n}$ is proportional to $b$. This corresponds to the ``direct emission'' scenario. The case of $n=2$ is for the ``lensing ring''. The $r_{\rm n}(b)$ function illustrates as a asymptotic curve, and $b$ is limited in a very narrow range around $b \simeq 4.73M-5.89M$ ($Q=0.5, a=0.5$). Therefore, the lensing ring shows up as a thin ring in the shadow image. The case of $n=3$ is for the ``photon ring''. In this scenario, $r_{\rm n}(b)$ is almost a vertical line, suggesting that the photon ring should be an extremely thin ring at a given parameter value. It is also found that the increase of the BH magnetic charge leads to a decrease of the impact parameter, the EH parameter do not influence it.

\par
\begin{figure*}[htbp]
  \centering
  \includegraphics[width=5.2cm,height=5.2cm]{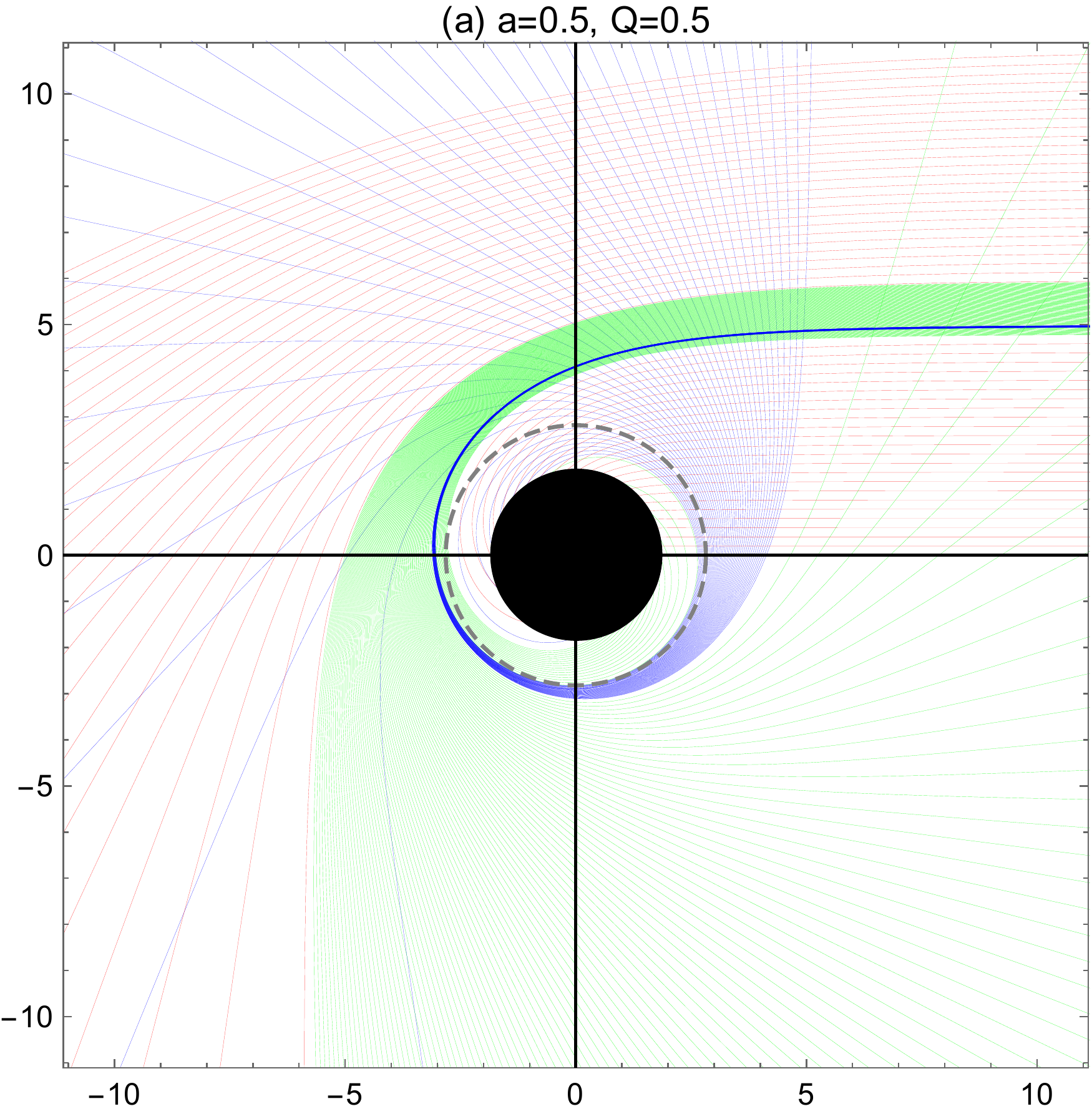}
  \hspace{0.5cm}
  \includegraphics[width=5.2cm,height=5.2cm]{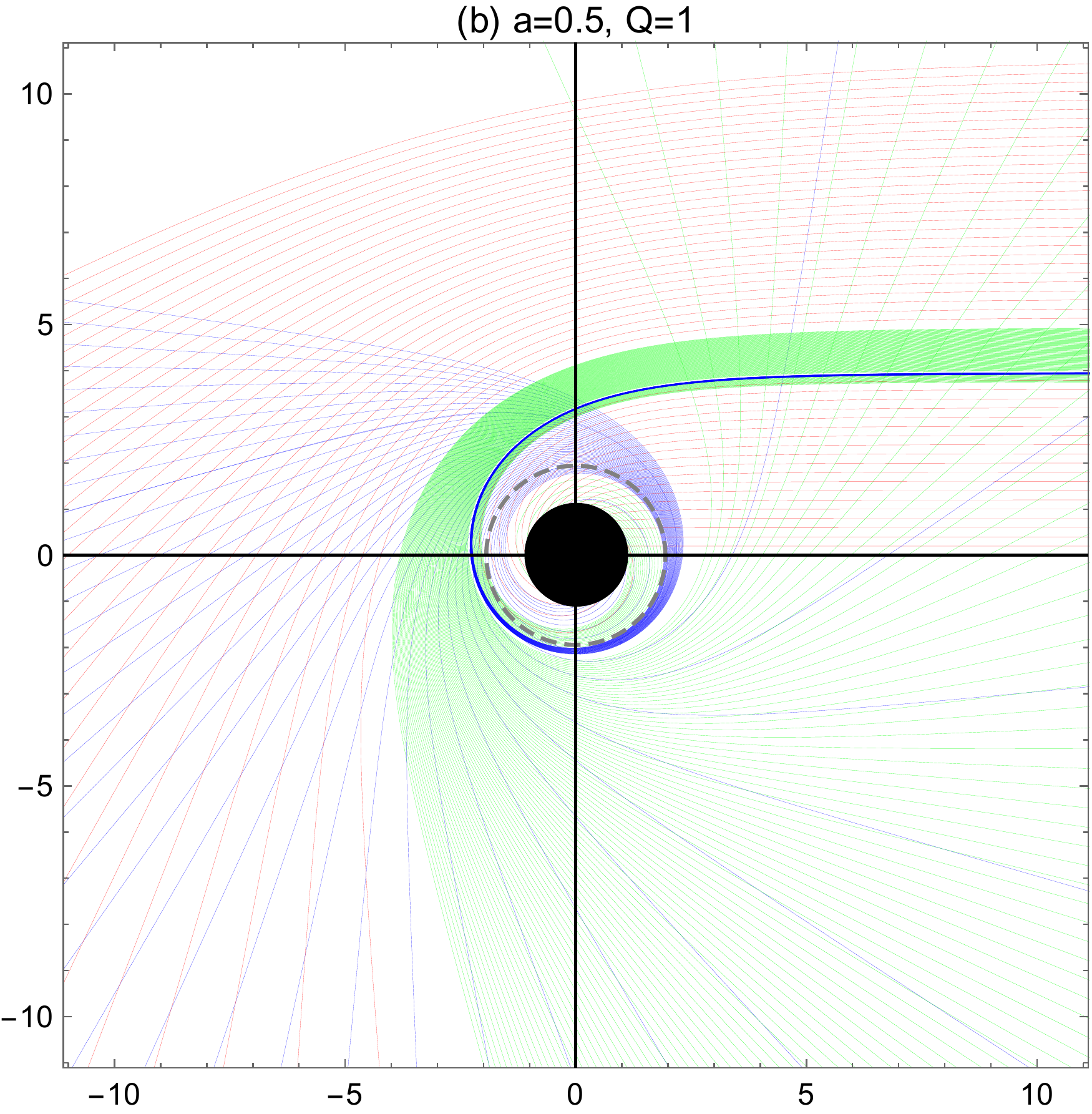}
  \hspace{0.5cm}
  \includegraphics[width=5.2cm,height=5.2cm]{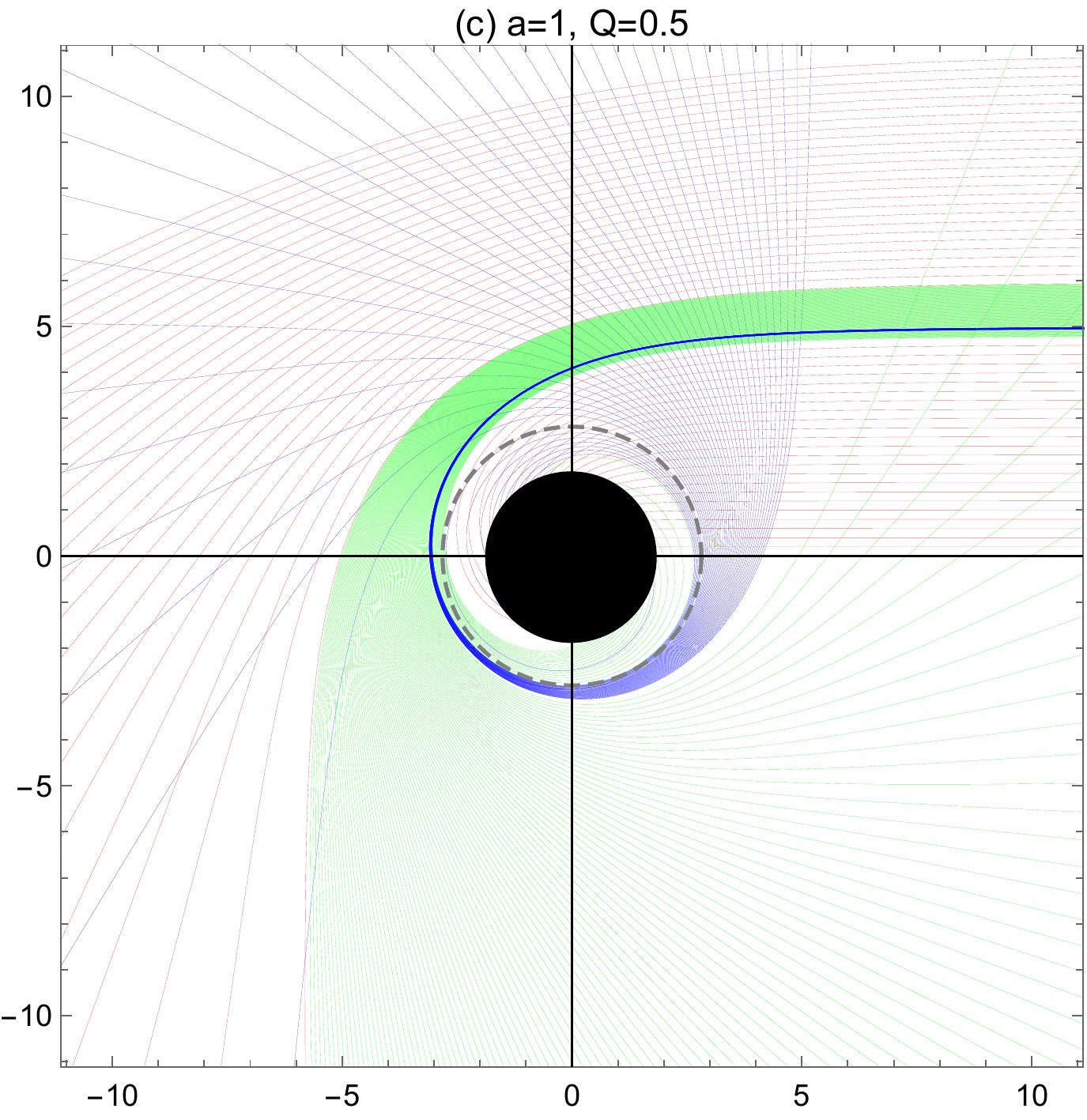}
  \caption {The light trajectories of different rings in the polar coordinate $(b,\phi)$. The BHs are shown as the black disks, and the dashed grey lines represent the BH photon ring orbits. The red lines, green lines, and blue lines correspond to the direct emission, lensing ring, photon ring, respectively. {\em Panel (a)}-- $a=0.5,~Q=0.5$, {\em Panel (b)}-- $a=0.5,~Q=1$, {\em Panel (c)}-- $a=1,~Q=0.5$. The BH mass is taken as $M=1$.}\label{fig:9}
\end{figure*}

Fig. \ref{fig:9} illustrates that the light trajectories of different rings in the polar coordinate by utilizing the ray-tracing procedure. One can see that the radius of the black disk is smaller and the light rays are more curved for the large BH magnetic charge. The radii of the direct emission, lensing ring, and photon ring are shrunk if the $Q$ value increases. Note that the thickness of the direct emission, lensing ring, and photon ring is not sensitive to the QED parameter.

\par
Next, we investigate the optical appearance of the EH BH in this scenario. We main purpose is to analyze the influence of the accretion disk radiation position on the observation characteristics of the BH. As well known that the radiation of accretion disk in the universe satisfies Gaussian distribution \cite{33}, hence, we parameterize the radiations intensity of the accretion disk as a gaussian function, that is
\begin{equation}
\label{3-3-5}
I_{\rm e}(r)~=~\left\{
\begin{array}{rcl}
exp\Big[{\frac{-(r-r_{\rm in})^{2}}{30}}\Big] ~~~~~~~~~~& & {r>r_{\rm in}},\\
0~~~~~~~~~~~~~~~~~~~~ & & {r\leq r_{\rm in}},\end{array} \right.
\end{equation}
where $r_{\rm in}$ is the innermost radiation position of the accretion disk. Our analysis is for three different scenarios: (A) $r_{\rm in}=r_{\rm isco}$, where $r_{\rm isco}$ is the radius of the BH innermost stable circular orbit; (B) $r_{\rm in}=r_{\rm ph}$ at which $r_{\rm ph}$ is the radius of the BH photon ring; (C) $r_{\rm in}=r_{+}$, where $r_{+}$ is the radius of the BH event horizon.

\subsubsection{Case A: $r_{\rm in}=r_{\rm isco}$}
\label{sec:3-3-1}
\par
\begin{figure*}[htbp]
  \centering
  \includegraphics[width=5.7cm,height=4.8cm]{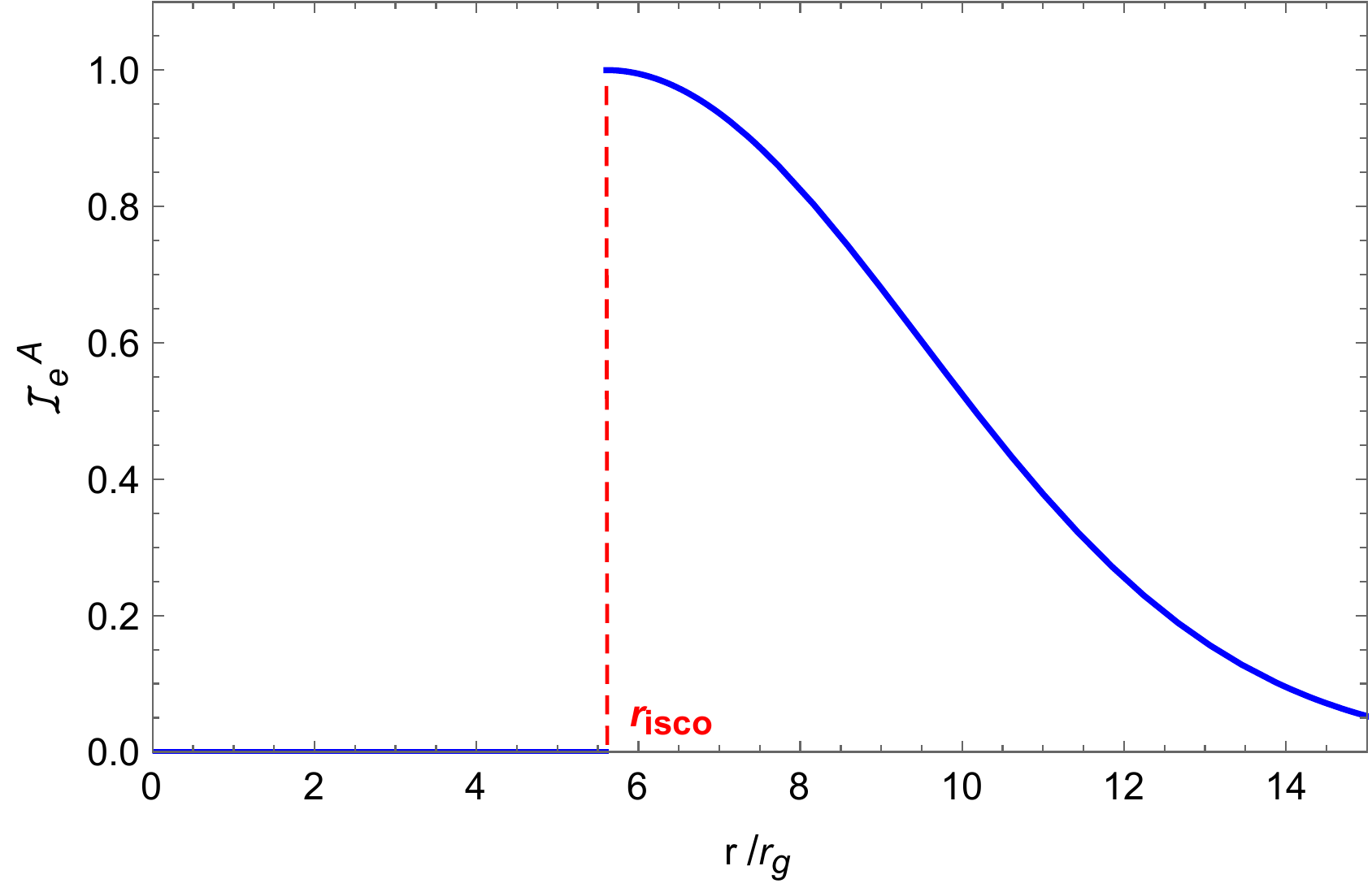}
  \includegraphics[width=5.7cm,height=4.8cm]{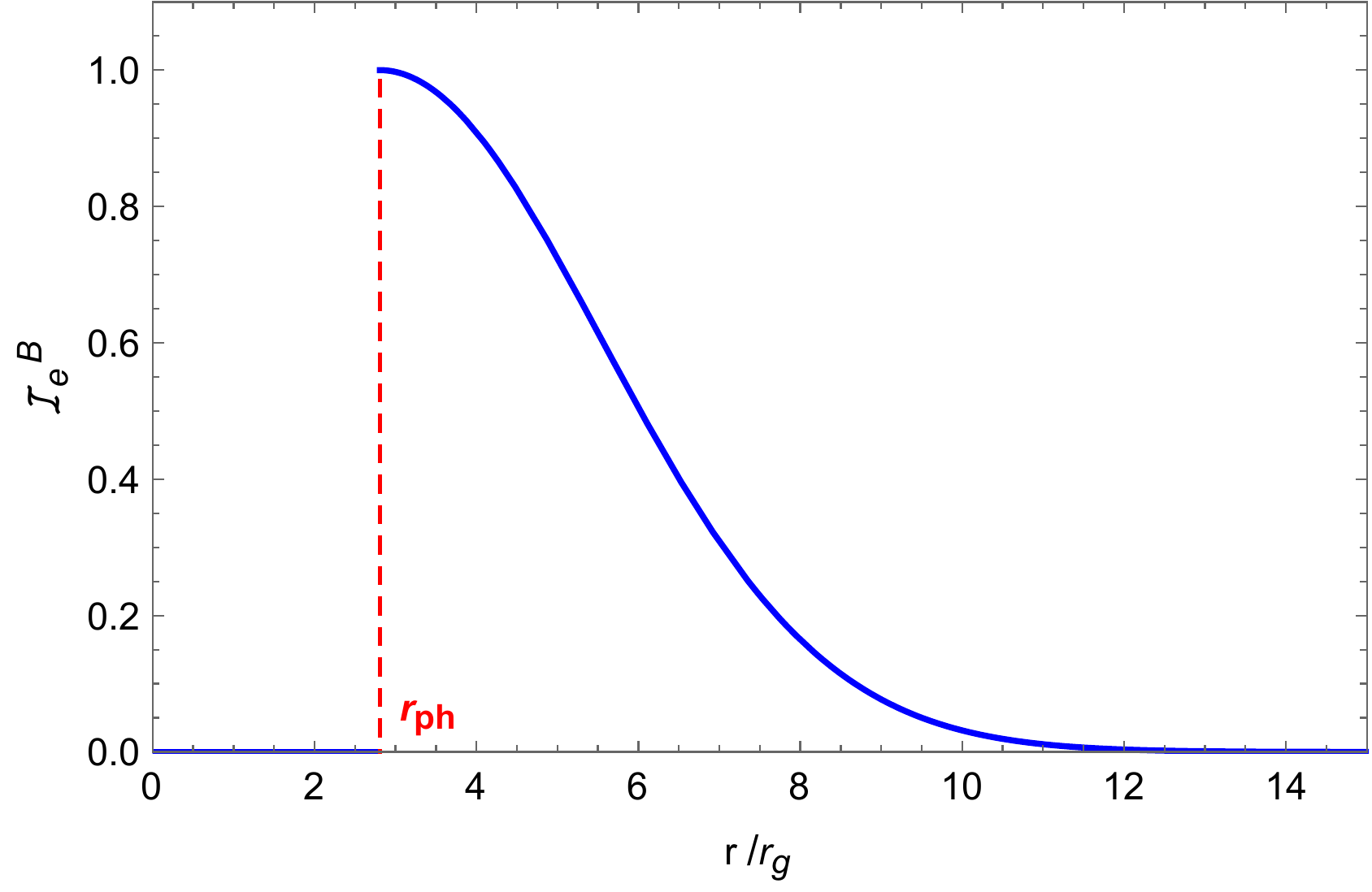}
  \includegraphics[width=5.7cm,height=4.8cm]{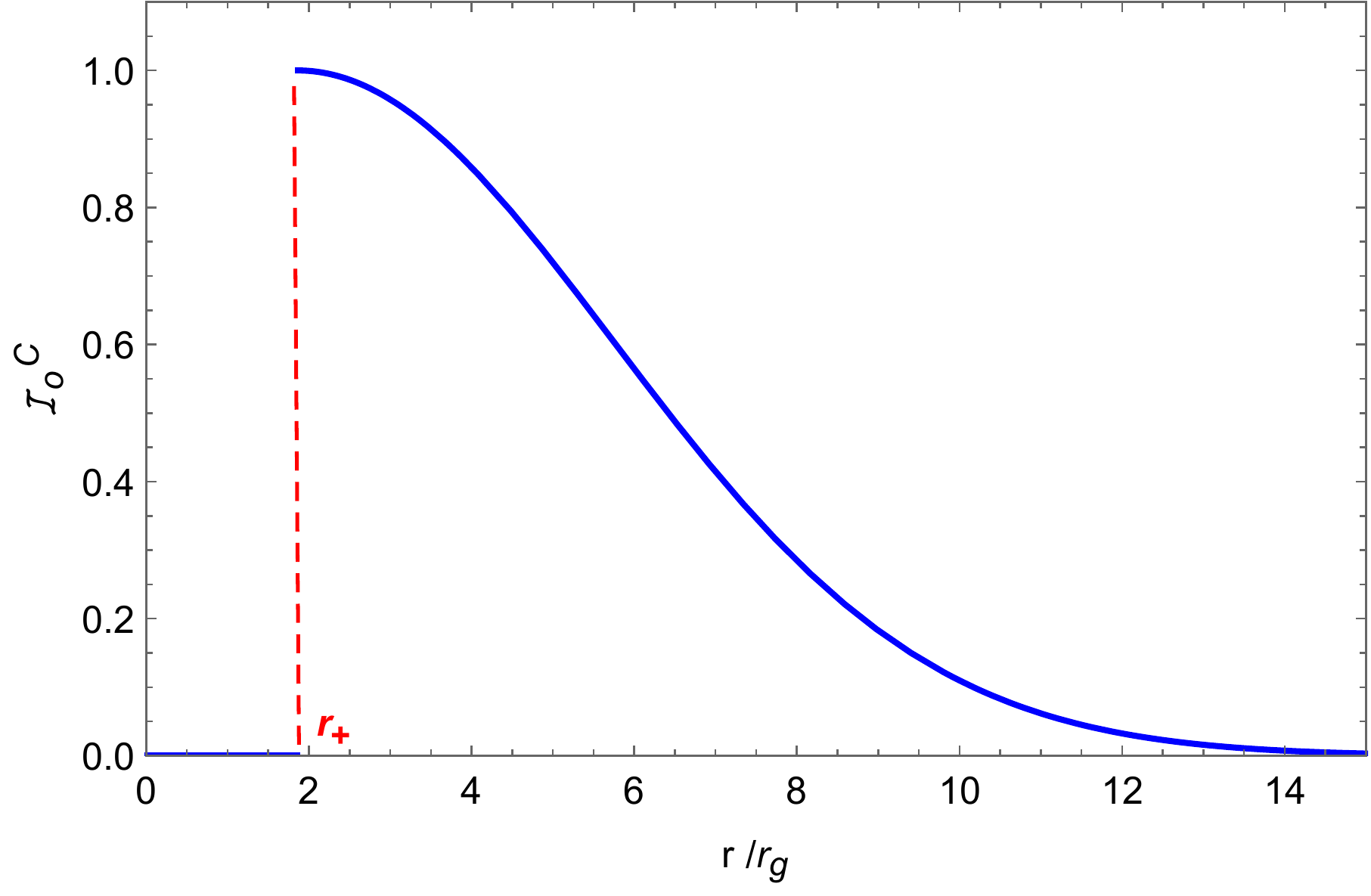}
  \includegraphics[width=5.7cm,height=4.8cm]{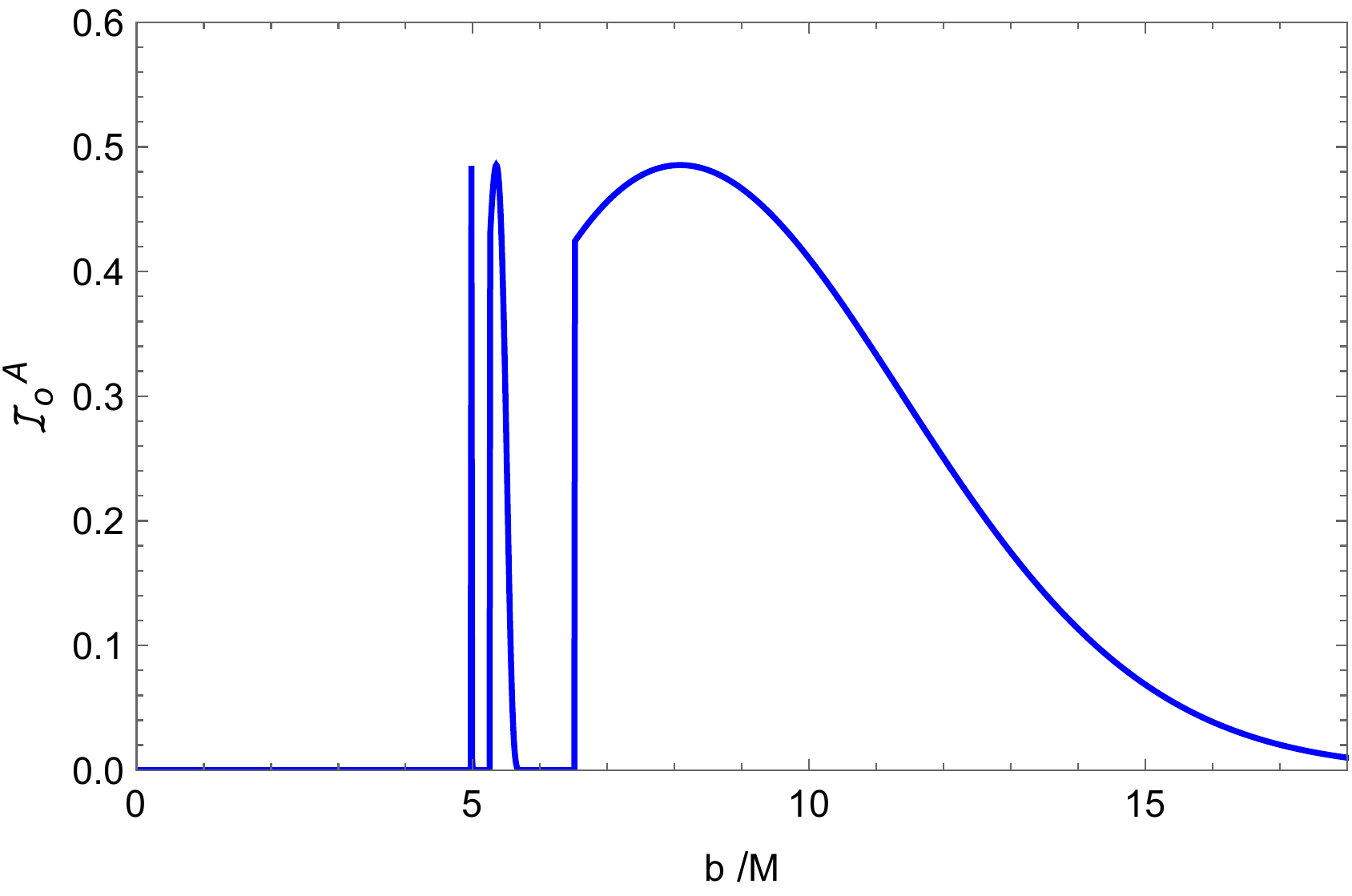}
  \includegraphics[width=5.7cm,height=4.8cm]{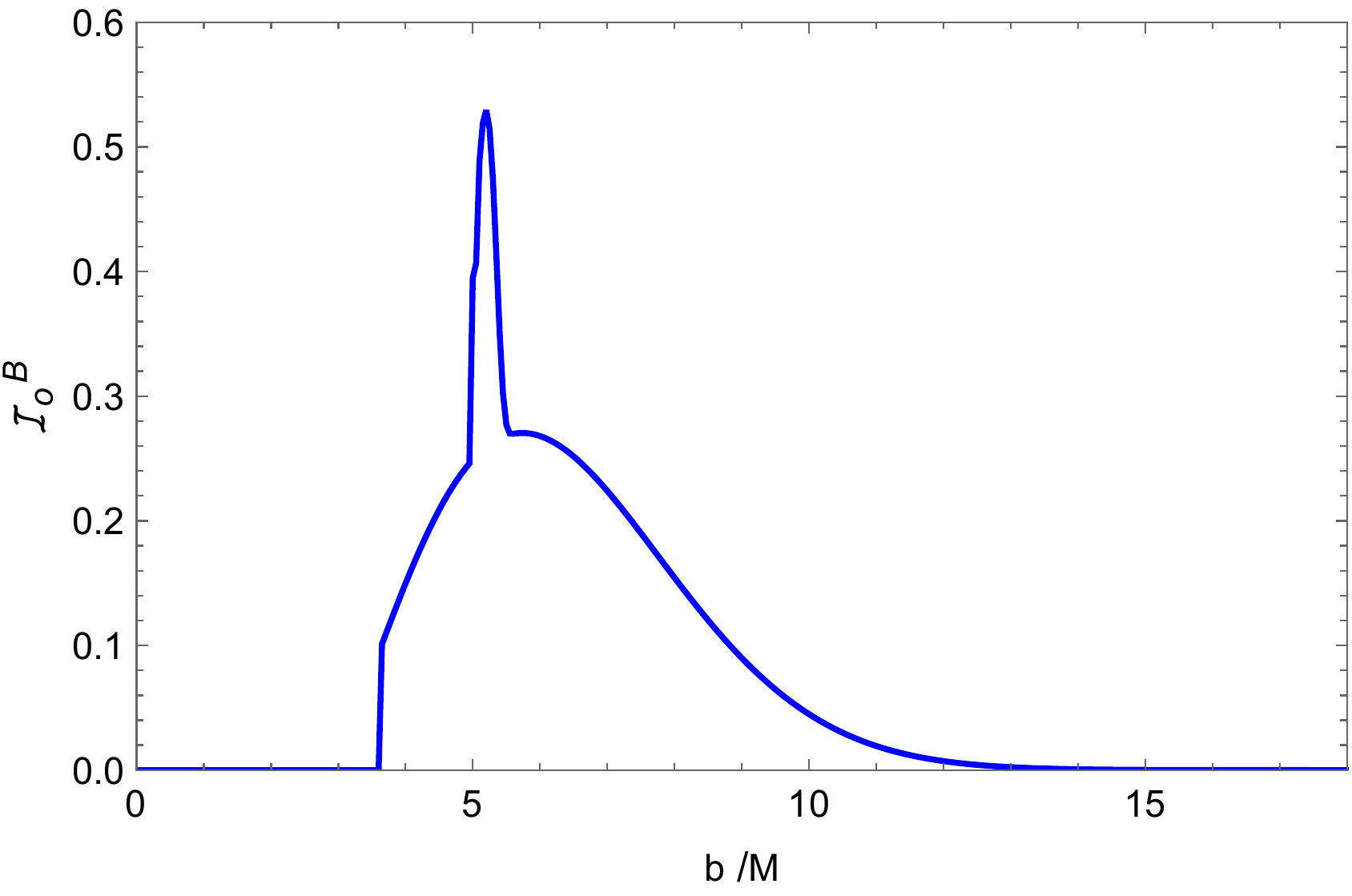}
  \includegraphics[width=5.7cm,height=4.8cm]{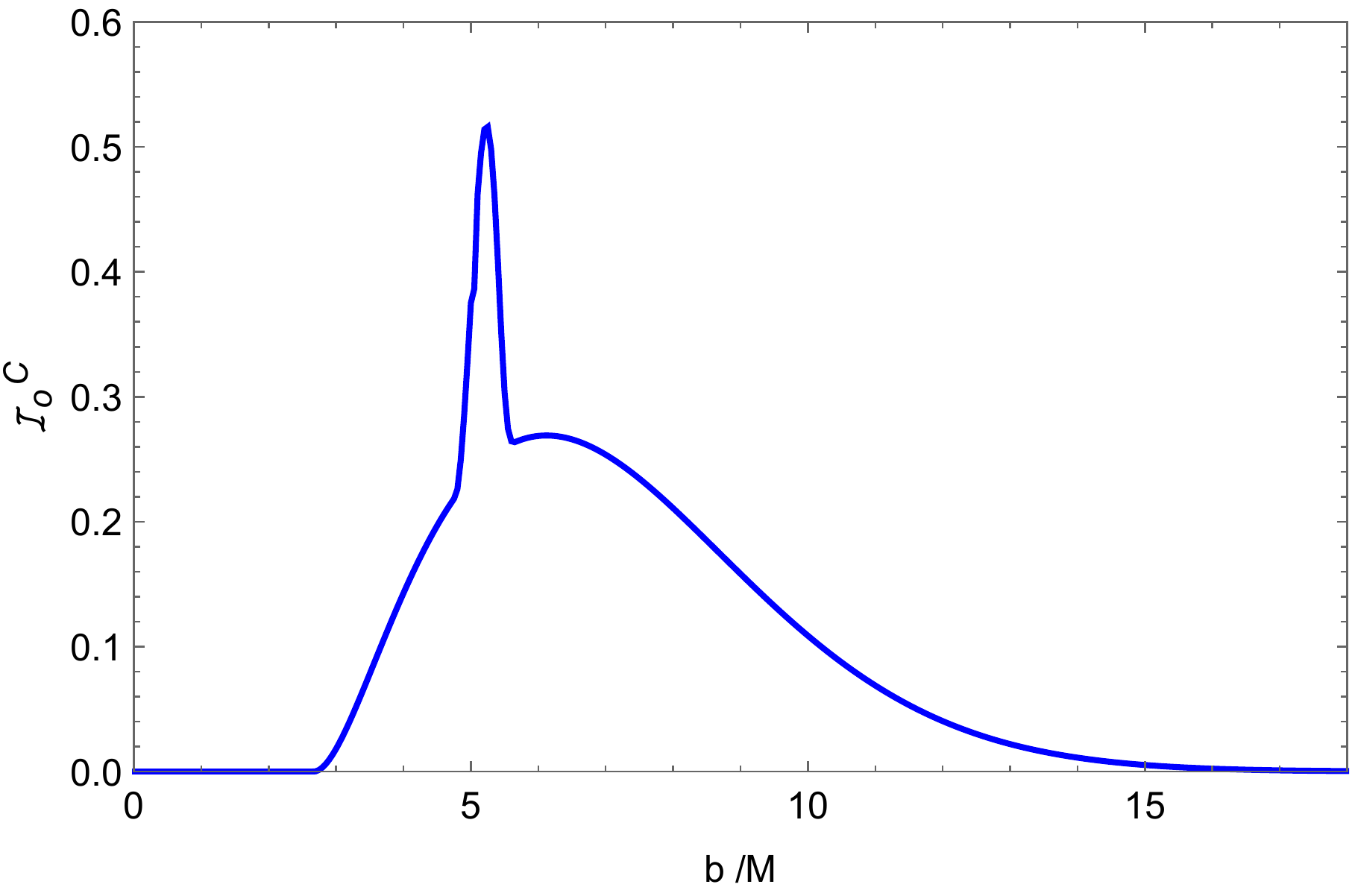}
  \includegraphics[width=5.2cm,height=5.2cm]{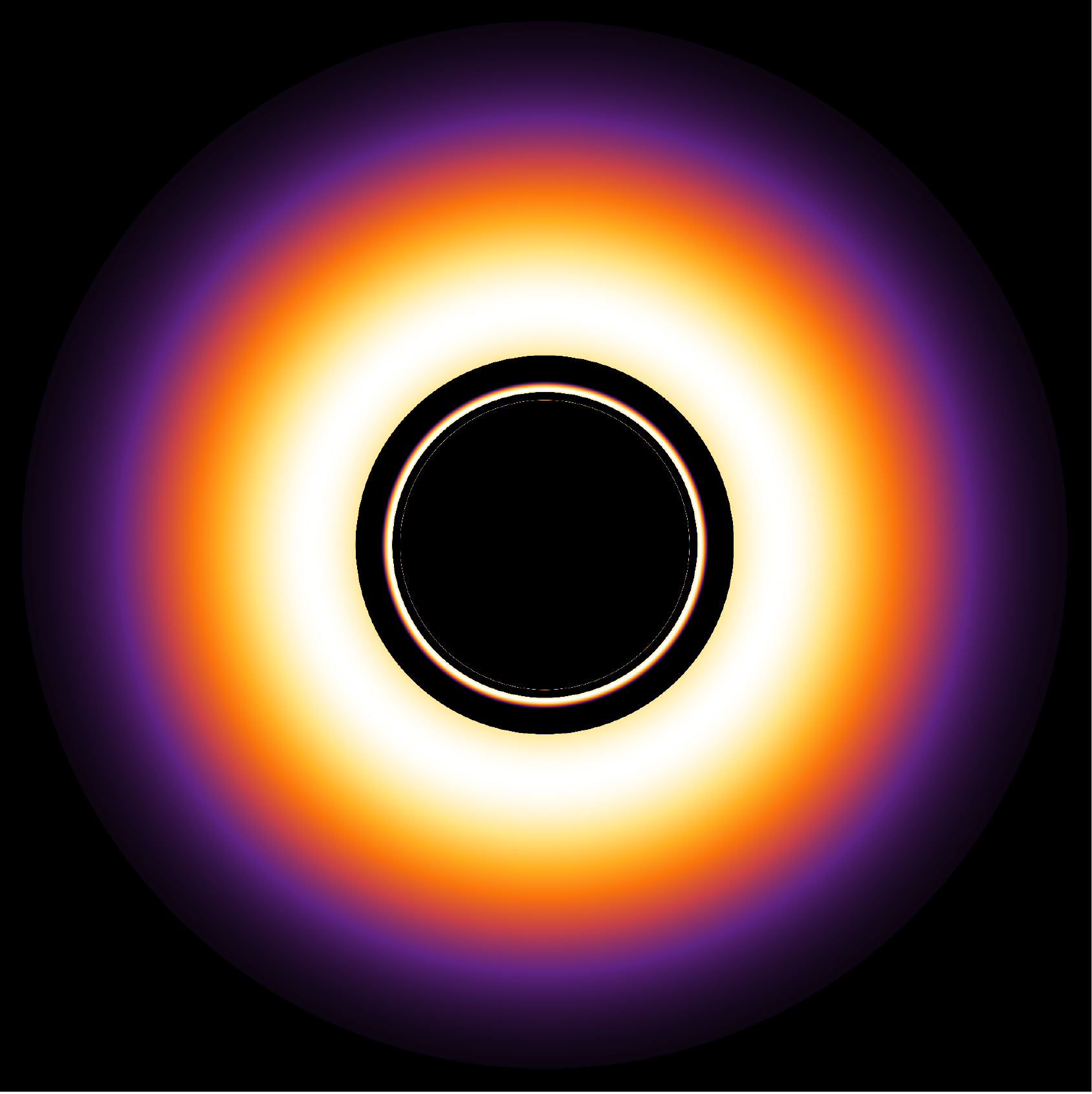}
    \hspace{0.35cm}
  \includegraphics[width=5.2cm,height=5.2cm]{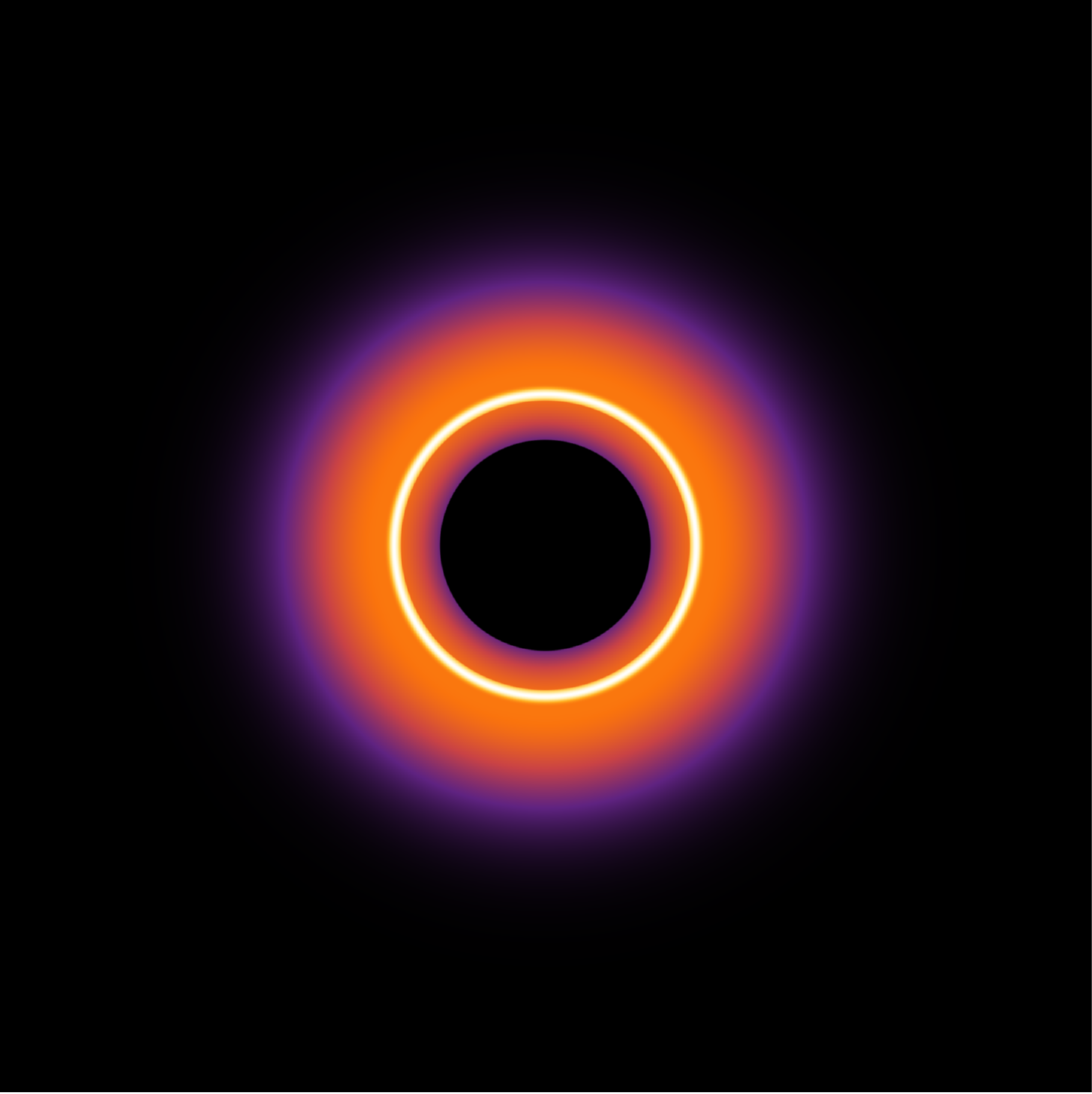}
    \hspace{0.35cm}
  \includegraphics[width=5.2cm,height=5.2cm]{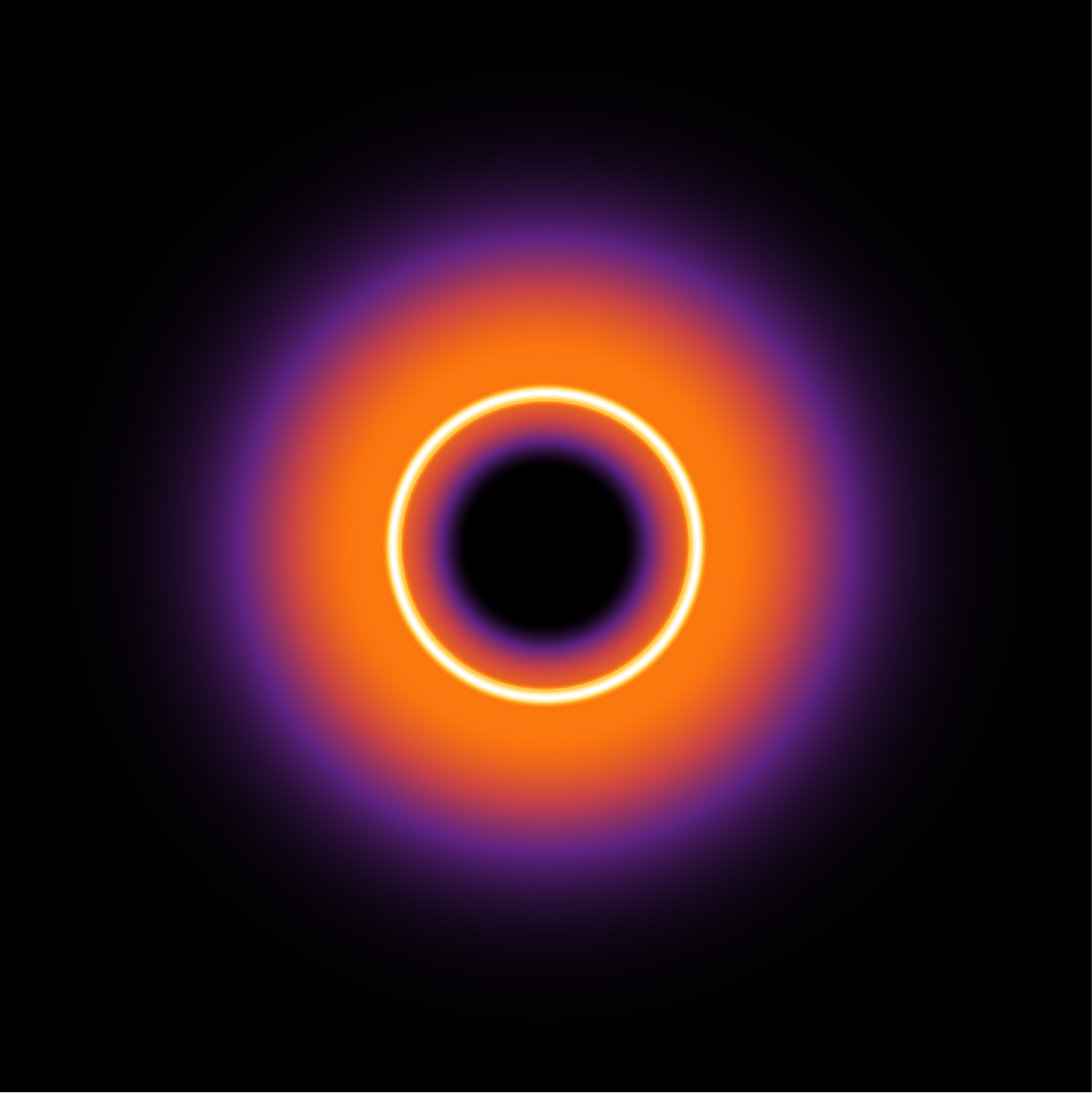}
  \caption {The total radiation intensity as a function of radius, the total observed intensity as a function of the impact parameter, and the two-dimensional images of the shadows for the EH BH. The $Q=0.5$, $a=0.5$, and the BH mass as $M=1$.}\label{fig:10}
\end{figure*}

The innermost stable circular orbit ($r_{\rm isco}$) is one of the relativistic effects, representing the bounder between test particles circling the BH and test particles falling into the BH. It is given by \cite{22}
\begin{equation}
\label{3-3-1-1}
r_{\rm isco}=\frac{3f(r_{\rm isco})f'(r_{\rm isco})}{2f'(r_{\rm isco})^{2}-f(r_{\rm isco})f''(r_{\rm isco})}.
\end{equation}
When the $Q=0.5$ and $a=0.5$, the innermost stable circular orbit radius of the EH BH is $r_{\rm isco}=5.61 r_{\rm g}$. According to Eqs. (\ref{3-3-4}) and (\ref{3-3-5}), the total radiation intensity $I^{\rm A}_{\rm e}$ as a function of the radius, the total observed intensity $I^{\rm A}_{\rm o}$ as a function of the impact parameter, and the two-dimensional image in celestial coordinates are displayed in the left panels of the Fig. \ref{fig:10}. It is found that the regions of the direct emission, lensing ring, and photon ring are separated. The direct emission starts at $b \simeq 6.52M$ and peaks at $b \simeq 8.18M$. Its maximum intensity is $0.485$. The lensing ring is limited in a small range of $b \simeq 5.24M \sim 5.65M$. 
The photon ring appears at $b \simeq 4.98M$. 
In the two-dimensional image, the border of the black disk corresponds to $r_{\rm isco}$. A bright lensing ring is shown within the black disk, and the dim photon ring is in the inner of the lensing ring.

\subsubsection{Case B: $r_{\rm in}=r_{\rm ph}$}
\label{sec:3-3-2}
Deriving $r_{\rm ph} \simeq 2.82 r_{\rm g}$ ($Q=0.5$ and $a=0.5$) from Eq. (\ref{1-17}). We show $I^{\rm B}_{\rm e}$ as a function of $r$, $I^{\rm B}_{\rm o}$ as a function of $b$, and the two-dimensional image in the middle panels of Fig. \ref{fig:11}. Different from Case A, the regions of the direct emission, lensing ring, and photon ring are overlapped. The direct emission starts at $b \simeq 3.62M$. Its maximum intensity is $0.25$. The lensing and photon rings are embedded in the direct emission region and limited in the range of $b\simeq 4.97M \sim 5.57M$. The photon ring appears at $b \simeq 5.21M$. The two rings are so close that it cannot be almost separated. If enlarge the two-dimensional image, one can observe that the photon ring clings to the lensing ring and presents a bright, extremely narrow ring. 
An completely dark region as shown in the BH photon ring.

\subsubsection{Case C: $r_{\rm in}=r_{\rm +}$}
\label{sec:3-3-3}
\par
When the innermost of the accretion disk radiation position is in the radius of the BH event horizon ($r_{\rm +}$), the radiation peak at the BH event horizon radius $r_{\rm +} \simeq 1.86 r_{\rm g}$. Our results are shown in the right panels of Fig. \ref{fig:10}. This scenario is similar to Case B. However, the black area that can be observed is reduced to the BH event horizon. The direct emission starts at $b \simeq 2.76M$ and the maximum intensity is $0.22$. The lensing ring is constrained to $b \simeq 4.78M \sim 5.62M$, and the photon ring generated at $b \simeq 5.24M$. 
It is also found that the positions of the lensing ring and photon ring is almost unchanged by compared with Case B. Meanwhile, the region of the photon ring of the BH becomes slightly wider since the attenuation of the accretion disk radiation in the third case is slightly slow.

\vskip 0.2cm
~~~~~~~~~~~~~~~~~~~~~~~~~~~~~~~~\emph{\textbf{Summary}}:
\vskip 0.2cm
\par
This section shows that the optical appearance of the EH BH shadow depends on the innermost radiation position of the accretion disk, which applies to the model of the BH surrounded by a thin accretion disk. In this scenario, the optical appearance of the BH shadows do not only depend on the structure of space-time, which is different from the static and infalling spherical accretion flow models. The contribution of the different rings to the total observed flux with different parameters are listed in Tab. \ref{Tab:3}. It is found that the contribution of the lensing ring to the total observed flux is less than $5\%$, and the photon ring is less than $2\%$, indicating that the direct emission dominates the optical appearance of the BH image.
\begin{table}[ht]
\caption{The total observed intensity corresponding to direct emission, lensed ring and photon ring of the EH BH with thin disk accretion flow, where the BH mass as $M=1$ and the magnetic charge taking as $g=0,0.2,0.4,0.6,0.8,1.0$.}\label{Tab:3}
\begin{center}
\setlength{\tabcolsep}{0.5mm}
\linespread{0.1cm}
\begin{tabular}[t]{|c|c|c|c|c|c|c|c|c|c|}
  \hline
  $g$/$Case$    &        $Case~ A$         &        $Case~ B$          &           $Case~ C$       \\
  \hline
  Em        &   $\rm Lensed;~~\rm Photon$   &   $\rm Lensed;~~\rm Photon$     &      $\rm Lensed;~~\rm Photon$    \\
  \hline
  $0$             &  $2.81\%;~~~1.04\%$   &   $3.56\%;~~~0.78\%$    &      $4.54\%;~~~1.25\%$    \\
   \hline
  $0.2$           &  $2.69\%;~~~0.81\%$   &   $3.47\%;~~~0.66\%$    &      $4.43\%;~~~1.17\%$    \\
   \hline
  $0.4$           &  $2.46\%;~~~0.68\%$   &   $3.35\%;~~~0.57\%$    &      $4.31\%;~~~1.06\%$    \\
   \hline
  $0.6$           &  $2.12\%;~~~0.52\%$   &   $3.02\%;~~~0.48\%$    &      $4.12\%;~~~0.84\%$    \\
   \hline
  $0.8$           &  $2.04\%;~~~0.39\%$   &   $2.96\%;~~~0.42\%$    &      $4.03\%;~~~0.71\%$    \\
  \hline
  $1.0$           &  $1.93\%;~~~0.28\%$   &   $2.88\%;~~~0.39\%$    &      $3.97\%;~~~0.62\%$    \\
   \hline
\end{tabular}
\end{center}
\end{table}

\par
\begin{figure*}[htbp]
  \centering
  \includegraphics[width=5.2cm,height=5.2cm]{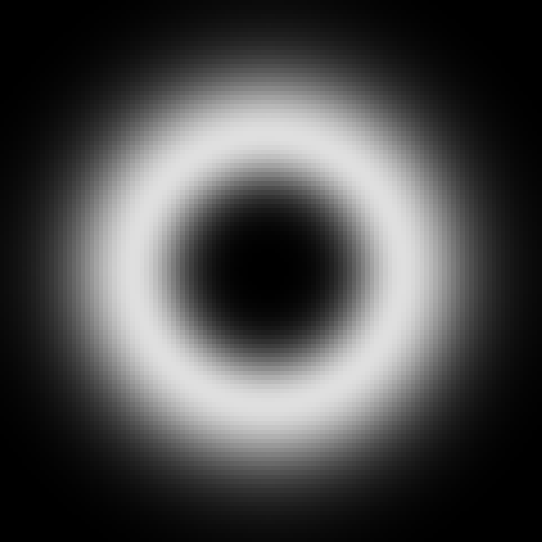}
  \hspace{0.5cm}
  \includegraphics[width=5.2cm,height=5.2cm]{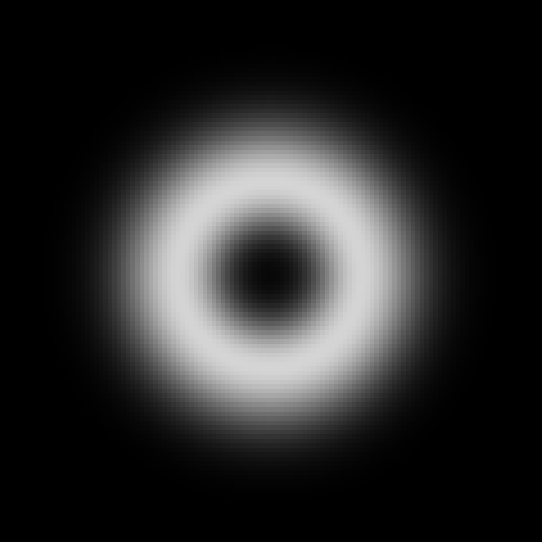}
  \hspace{0.5cm}
  \includegraphics[width=5.2cm,height=5.2cm]{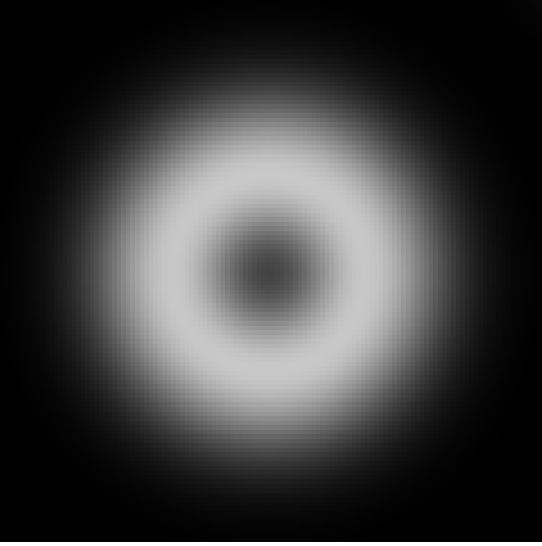}
  \caption {Blurred two-dimensional images utilizing a Gaussian filter with a standard deviation of 1/12 the field of view for the thin disk accretion flow in three cases.}\label{fig:11}
\end{figure*}

We also blur the two-dimensional images as shown in Fig. \ref{fig:11}. One can observe that the blurring washes out the lensing ring and photon ring features. Their observational appearances reply on instrument resolution. It is difficult to obtain the ring information with the current resolution of EHT. Thus, we suggest that the optical appearance of the BH image depend on the accretion disk radiation position in this scenario.

\section{\textbf{Conclusions and Discussion}}
\label{sec:4}
\par
The optical appearance of the EH BH in the different accretion flows framework have been revealed in this analysis. Taking the QED effect into account, we obtained that the effective geometry induced by the EH non-linear electrodynamics effects. The photons propagate along the null geodesics for this effective geometry is discussed. According to derived the effective potential function, we found that the increase of the magnetic charge leads to the decrease of the EH BH event horizon radius, shadow radius and critical impact parameter, implying that the BH photon ring is shrunk inward the BH by increasing the magnetic charge. By investigating trajectory of the light ray, we found the radius of the black disk is smaller and the light rays are more gentle near the BH for a larger magnetic charge and the light density increases with the increase of magnetic charge.

\par
For the EH BH is illuminated by the static and infalling spherical accretion flows, the total observed intensity function is calculated. We found that the observed intensity first ascended with the impact parameter, and reached the peak at the BH photon ring. The peak value of intensity increase with an increase of the BH magnetic charge, and the corresponding $b_{\rm ph}$ get smaller. The change of $a$ does not affect the size and luminosity of the shadow, which indicates that the luminosity of the EH BH is independent of the one-loop corrections to QED effect. We also found the total observed intensity in the static spherical accretion flow scenario leads than that of the infalling spherical accretion flow under same parameters. The size and position of the EH BH shadows do not change in both of these accretion flows, implying that the BH shadow size depends on the geometric space-time and the shadows luminosities relies on the accretion flow models.

\par
For the EH BH is surrounded by a thin disk accretion flow, we found that the radii of the direct emission, lensing ring, and photon ring are dramatically decreased as the magnetic charge increases. In case of that the innermost of the accretion disk radiation position equals to $r_{\rm isco}$, the regions of the direct emission, lensing ring, and photon ring are separated. A bright lensing ring is shown within the $r_{\rm isco}$, and the dim photon ring is in the inner of the lensing ring. In case of the innermost of the accretion disk radiation position equals to $r_{\rm ph}$, the two rings are so close that it cannot be almost separated. In case of the innermost of the accretion disk radiation position shrinks to the event horizon of the BH, the observable black area is shrunk to the BH event horizon, but the radii of the lensing ring and photon ring is almost unchanged. We also found that the contribution of the lensing ring to the total observed flux is less than $5\%$, and the photon ring is less than $2\%$, indicating that the direct emission dominates the optical appearance of the BH image. Blurring the two-dimensional image in this scenario, we observed the blurring washes out the lensing ring and photon ring features, which replies on instrument resolution. It is difficult to obtain the ring information with the current resolution of EHT. Thus, we believe that the optical appearance of the BH image depend on the accretion disk radiation position.

\par
Based on our analysis, we argue that the optical appearance of the EH BH depends on the accretion flows morphology and BH space-time structure. When the spherical accretion flows illuminated the BH, the BH shadow is shown as a geometric feature of space-time, the size of the BH shadow does not change with the position of the spherical accretion flows. When the thin disk accretion flow illuminated the BH, the observable characteristics of the BH shadow rely to the position of the radiating accretion disk with respect to the BH.

\setlength{\parindent}{0pt}\textbf{\textbf{Acknowledgments}}
This work is supported by the National Natural Science Foundation of China (Grant No. 12133003, 11851304, U1731239, 11675140, 11705005, 11903025 and 11875095).\\

\end{CJK}
\end{document}